\documentclass[12pt]{cernart}
\usepackage{epsfig}
\usepackage{amssymb}

\def\be{\begin{equation}}
\def\ee{\end{equation}}
\def\bea{\begin{eqnarray}}
\def\eea{\end{eqnarray}}

\def\beq{\begin{equation}}
\def\eeq{\end{equation}}

\begin{document}
%%%%%%%%\psdraft %non mette le figure
%%%%%%%%%%%%%%%%%%%%%%%%%%%%%%% titlepage %%%%%%%%%%%%%%%%%%%%%%%%%%%%%%%%%%%%
\begin{titlepage}
  \docnum{CERN--EP/99--126}
  \date{24 August 1999}
  \title{ Inferring the intensity of Poisson processes \\
         at the limit of the detector sensitivity\\
         (with a case study 
         on gravitational wave burst search)}
         \author{P.Astone\Instref{xx} and G.D'Agostini\Instref{yy}}
% Submitted{Annals of Physics}
% collaboration{}
% conference{}
% note{}
% dedication{}
\Instfoot{xx}{Sezione INFN di Roma 1, Rome, Italy\\
   {\rm Email}: {\tt astone@roma1.infn.it} \\ 
   {\rm URL}: {\tt http://grwav3.roma1.infn.it}}
\Instfoot{yy}{ Universit{\`a} ``La Sapienza'' and
        Sezione INFN di Roma 1, Rome , Italy, and CERN, Geneva, Switzerland\\
        {\rm Email}: {\tt dagostini@roma1.infn.it}\\ 
        {\rm URL}: {\tt http://www-zeus.roma1.infn.it/$^\sim$agostini/}}
%
%\begin{titlepage}
%\begin{flushright}
%        \small
%          hep-ph/9909047\\
%         Roma1-10XX\\
%         March 1999
%\end{flushright}
%
%{\large\bf Abstract}
%
%\vspace{.5cm} 
%\end{center}
\begin{abstract}
We consider the issue of reporting 
the result of search experiment in the most unbiased and
efficient way, i.e. in a way which allows an easy
interpretation and combination of results and which
do not depend on whether the experimenters believe or
not to having found the searched-for effect. 
Since this work uses the language of Bayesian theory,
to which most physicists are not used, we find that it could
be useful to practitioners to have in a single paper
a simple presentation of Bayesian inference, together
with an example of application of it in search of rare processes.
\noindent
\end{abstract}
\vspace{7cm}
\submitted{(Submitted to Annals of Physics)}
\end{titlepage}
%%%%%%%%%%%%%%%%%%%%%%%%%%%%%%%%%%%%%%%%%%%%%%%%%%%%%%%%%%%%%%%%%%%%%%%%%%%%%
%%%%%%%%%%%%%%%%%%%%%%%%%%%%%%%%%%%%%%%%%%%%%%%%%%%%%%%%%%%%%%%%%%%%%%%%%%%%%
% from the dvips manual: put a background `DRAFT' on the page
%\special{!userdict begin 
%/bop-hook{gsave 200 30 translate 65 rotate
%           /Times-Roman findfont 216 scalefont setfont
%           0 0 moveto 0.95 setgray (DRAFT) show grestore}def end}
%
%%%%%%%%%%%%%%%%%%%%%%%%%%%%%%%%%%%%%%%%%%%%%%%%%%%%%%%%%%%%%%%%%%%%%%%%%%%%%
\setcounter{page}{2}
\section{Introduction}
An often debated issue\footnote{The many recent 
papers~\cite{Janot1,Favara,FC,PDG,Janot2,Giunti1,Giunti2,
Zech,Jim,Roe,Giunti3,Junk,Yellin,Geer,Narsky,Hu,Helene} on `limits', 
not to mention notes internal to  
experimental teams, 
 give an idea of the present interest 
in the subject. However, this article is not 
a review of the various `prescriptions' suggested 
by the many authors involved in the discussion.
In fact, the point of view presented in this paper is that
the search of the Holy Grail containing the unique 
and objective prescription to calculate limits is a false problem. 
Since the cited papers --- with the exception
of Ref.~\cite{Zech} and, to some extent, Ref.~\cite{Helene} --- 
are written in this spirit, they are irrelevant for 
this work. Another common point of all the cited authors, 
with the sole exception of Ref.~\cite{Zech}, is to 
consider the frequentist 
concept of coverage as good guidance. Zech~\cite{Zech} 
considers coverage 
{\it ``a magic objective of classical confidence bounds. It has
an attractive property from a purely 
aesthetic point of view but it is not obvious how 
to make use of this concept''}. Our opinion
about frequentistic coverage is even more severe, 
and it has been discussed extensively in Ref.~\cite{dagocern}.
Moreover, comments on Refs. \cite{FC} and \cite{PDG}, which 
have triggered most of the cited papers, can be 
found in Refs. \cite{maxent98} and \cite{priors}. 
In particular, one should not overlook the fact that the
results obtained by frequentistic confidence intervals, as well 
as those obtained by frequentistic hypothesis tests, are usually  
misunderstood and might even induce researchers to 
draw misleading scientific 
conclusions~\cite{maxent98,dagocern}.} 
in frontier science research is how to
report results obtained from search experiments at the limit 
of the detector sensitivity. Sometimes researchers
have simply to state a clear null result, i.e. when all 
members of the experimental team agree that no new phenomenon is 
indicated by the data. 
At other times they may have some hints
that the data could indicate the presence of the searched-for
signal, as a result of 
a more or less pronounced excess of events
above the expected background level. In lucky, and rare, cases 
 new phenomena are seen in such a spectacular 
way that all researchers agree and everybody is convinced. 
Clearly, reporting the result may become problematic 
in the second case. 
 ``The experiment was inconclusive, 
and we had to use statistics'', somebody once said. 

The purpose of this paper is to show how results of 
the search for rare phenomena can be presented, 
in order to best use the information 
contained in  the experimental data, i.e. 
in the most powerful and unbiased way. Since
the three situations sketched out above are, in reality, never 
so sharply separated, the presentation of the result should not
depend on whether researchers feel that their case is a      
negative, doubtful, or positive one. 
Moreover, it is important that the pieces of evidence
from different experiments can be combined in the most
efficient way.   
If, for example, many independent data sets 
each provide a little evidence in favour of the searched-for 
signal, the combination of all data should enhance that 
hypothesis. If, instead, the indications
provided by the different data sets are incoherent, 
their combination  should provide a stronger constraint on the
intensity of the postulated process. 

Typical fields of research in which the above described 
 problematic situation arises are, to give a few examples,
neutrino oscillations, rare decays, new particles,  
gravitational waves, and dark matter. 
All these processes have in common the fact that, under 
stationarity of the search conditions, the physical 
process can be modelled with high accuracy by a Poisson process, 
and the physical quantity (a mass, a cross-section, 
a branching ratio, a rate, etc.)
of interest will be related 
to the intensity $r$ of that process. 
%The physical quantity 
%could be a branching ratio, the mass of a postulated particle, or
%the scale of a new interaction.

Although the 
methods described in this paper are of general use,
we think that they  can be better understood by way of 
a case study. We consider the problem of inferring 
the rate of bursts of gravitational waves (g.w.) 
on Earth.
This case presents 
typical features  common to other frontier searches,
but the problem remains unidimensional, since only one
quantity is inferred,\footnote{For example, 
in neutrino oscillation search the results 
are given in terms of 
the mixing angle and of the mass-squared difference.
Note that, also in this case, it would be very interesting 
to have the result in terms of cross-section of the process
searched for, kept separate
from the interpretation in terms of the postulated oscillations.
Then, one would deal also in this case with unidimensional problems,
i.e. cross-sections for bins of the incoming neutrino energy.}
and thus easy to describe. The extension to 
higher dimensions is, at least conceptually, straightforward. 

Since the methods used in the paper are based on 
 Bayesian inference  and subjective probability, 
to which most researchers are at present not accustomed, 
we feel that it is necessary
to introduce this matter in an extensive and elementary way. 
In particular, we also think it is important to clarify 
some of the philosophical aspects which physicists tend to ignore,
but which are crucial to the understanding and acceptance of 
the inferential framework which will be used.
Therefore, we consider it is convenient for the reader to have 
a description of the case study
and of the inferential framework in an almost self-contained
article.\footnote{Brief and extensive physicist's 
introductions to subjective probability and Bayesian inference 
can be found in Refs.~\cite{dagocern} and \cite{ajp}, respectively.} 
 
The paper is structured in the following way. In the next
section we recall the present status and future prospects
of g.w. burst search, stressing some of the important
aspects of the experiments which affect our 
general considerations
about the analysis strategy.
Then, the inferential
framework to be used to report results will be 
presented and discussed
in depth, though remaining at an introductory level. 
In the core of the paper the inferential model will be applied 
to the case study, with general considerations
and numerical examples.
Finally some conclusions will be drawn.  

\section{Gravitational-wave burst search}
\subsection{Status and perspective}
The interest in g.w.'s is related 
to the astrophysical information they contain 
and also to the implications their detection would have for 
fundamental physics~\cite{schutzwag}. 
Their detection would, in fact, lead to confirmation
of  Einstein's general relativity predictions 
in a more direct way than Hulse and Taylor's
observations~\cite{hulse}.
Gravitational-wave detectors could provide a direct 
measure of the waves and
could also test their properties. In particular,
 using a network
of detectors, wave speed  and polarization state 
can be inferred. Regarding
the emission process, the importance 
of g.w.'s lies in 
the fact that they pass through matter without being
significantly absorbed or scattered, unlike 
electromagnetic waves and even  weakly interacting neutrinos.
Thus the information about the emission process 
carried by g.w.'s is really unique 
(see Ref.~\cite{thorne} for a review of g.w. sources).

At present, one of the most interesting  activities 
within that section of the community which is operating resonant 
antennae is the search 
for evidence of g.w.  bursts.  
These are defined as bunches of g.w.'s 
whose time width
is smaller than the time constant of the detectors
(the latter being typically of the order of milliseconds).
Thus their
energy spread is expected to be flat across the whole of the 
frequency bandwidth of the detector.
A burst of g.w.'s can be produced in a gravitational collapse
associated with supernova explosions, or during the final stage of the
coalescence of binary systems (neutron stars, black holes), or in
processes involving massive black holes, such as the capture
of a near body.

Astrophysical estimates of rates and signal
amplitudes for these processes on Earth would seem discouraging
in the light of the present theoretical ideas.
In fact, given the sensitivity of the present antennae,
the expected rates are
much too low to give a sizable excess of candidate events
above the expected background.
The available detectors could, in fact, detect a g.w. burst
from the Galaxy if a process radiated
$1 \%$ of a solar mass ($M_\odot$), which would yield
a dimensionless wave amplitude on
Earth of $10^ {-18}$~\cite{pizz,bardonecchia}.
However, the expected supernova rate in the Galaxy could be
in the range of one per 10-100 years (see e.g. Ref.~\cite{blair})
and an emission of $1 \%$ $M_\odot$ into g.w.'s seems quite
improbable. Nevertheless, there is no solid ground for supposing that 
this hypothesized
fraction of energy is released into g.w.'s and even larger
fractions are conceivable~\cite{schutzwag}. 
Moreover, the burst rate could increase by a factor of about $1000$
if the antennae were sensitive to astronomical events
within a distance of 10 Mpc, thus including the Virgo cluster.
Improved bar detectors~\cite{nautilus, auriga}, as well as
planned interferometers~\cite{schutz},
are expected to reach this level of sensitivity.

In conclusion, although current prospects are not encouraging,
the many uncertainties on the physics
processes involved might still mean that surprises are in store 
and for this reason
it is important to be prepared to exploit to the full the information
provided by operating and planned detectors.

\subsection{Search strategy} 
Gravitational-wave bursts are very weak signals,
embedded in the noise of the detector.
Thus, they can be extracted from the detector data
by proper filtering, optimized to increase
the signal-to-noise ratio (SNR) for this class of events
\cite{filtrof}. 
The analysis is very
difficult because of the  low SNR, the rarity of the events,
the uncertainty regarding their shape, and the non-stationary
noise of the detectors \cite{sergio}.
In fact, although some of the sources of noise, like 
narrow-band Brownian noise and  electronic 
wide-band noise, are well understood and their expectations can be 
modelled with reasonable accuracy, there are 
other sources of background
which are not easy to handle, and not even easy to recognize.  

A filter for g.w. burst search 
is optimized to increase the SNR for $\delta$-like
signals, and a candidate event is defined when the 
filtered signal exceeds a certain energy 
threshold. The candidate event
is characterized by energy, arrival time, above-threshold 
duration, and other relevant quantities related to the
spectral content in different bandwidths \cite{sergio}.
All  event characteristics are, in fact, important. 
For example, the shape of the electric 
signal coming from the transductor can be used
to discriminate g.w. bursts from background. 

The rarity of the events looked
for and the presence of irreducible background
make it impossible to do this search using a single
detector, even if seismic, electromagnetic and 
other sensors are often used
 to veto the data
of a g.w. detector (see e.g. Refs. \cite{explorer,nautilus}).
Therefore, a coincidence among at least two parallel and distant
detectors is
required.\footnote{`Parallel' means oriented 
in such a way as to be
sensitive to the
same polarization and direction of the 
incoming wave, and `far' means
located at a distance such that the 
long-range correlated background
is considered to be negligible.} 
Gravitational-wave bursts are, in fact, supposed 
to irradiate the Earth uniformly
so that detectors spread out across the Earth's surface
should be able to detect g.w.'s related to 
the same physical event.
Hence the whole analysis procedure consists 
of data filtering, event
selection, vetoes when necessary, and  
the final coincidence analysis.

An important parameter of the procedure
for extracting g.w. burst candidates
is the coincidence window, that is the time width 
within which the coincidences are considered.
The window can be fixed by considering the physics of the process
and the characteristics of the apparatus~\cite{nostro2}.

At present five resonant g.w. antennae are in operation, and this
is really the first time that it is possible to
search for g.w.'s 
with such a high number of detectors working
simultaneously:
Explorer\,\cite{explorer}, NAUTILUS~\cite{nautilus} 
and AURIGA~\cite{auriga} in Italy; Allegro\,\cite{allegro} in USA;
Niobe\,\cite{niobe} in Australia.
A collaboration has been established between the 
experimental groups, with the aim of 
performing coincidence searches of g.w. bursts. 
The joint effort resulted in 1997  
in a data exchange  protocol, in which 
candidate events are precisely defined and the procedure
for exchanging data was agreed 
(see e.g. Ref.~\cite{igec} and related web  sites).
It is, then, important to agree on an optimal way for publishing
results such that all information contained in the data
can be used in the most efficient way. 

Coincidence experiment procedures are essentially  the same as those 
used since the beginning
of g.w. experiments~\cite{weber}. Recently they 
have been used for
the analysis of Explorer and Allegro 1991 data~\cite{bardonecchia} 
and   analyses of 
Explorer-NAUTILUS (1994-1996) and Explorer-Niobe (1995)~\cite{coinci}.
The only relevant background to coincidence
analysis is due to accidental coincidences, which can be
estimated with high accuracy by off-timing techniques.

For the sake of simplicity we consider here only 
 coincidences between a couple
of parallel detectors. 
The rate of background ($r_b$) due to the 
accidental coincidences 
between the candidate events
is usually evaluated by the average
of the coincidences at shifted times.
Alternatively, one can make use of individual background 
rates ($r_{i_1}$ and $r_{i_2}$) and of the coincidence window 
$w$ to evaluate $r_b$ as $r_b=r_{i_1} r_{i_2} w$.
The two estimations of the expected accidental coincidence rate 
usually give the same result~\cite{nostro2}.
Once $r_b$ is evaluated, the observed number of  coincidences
in a given observation time $T$
due to background is described by a Poisson distribution. This is  
because accidental coincidences fulfil the conditions
which define a Poisson process, if the noise is stationary during $T$.
Therefore, the observed 
frequency distribution of off-timing coincidences is expected 
to be very close to the Poisson probability distribution of parameter
$\lambda_b=r_b\,T$.
As the distributions actually 
observed are indeed of that kind,  researchers are highly confident 
about the probability distribution of background coincidences.

\section{Probability of accidental coincidences versus 
probability of burst rates}
Let us begin by illustrating the kind of problems that can arise
in interpreting results of   
coincidence experiments, if not properly stated.  
Let us imagine that
$n_c$ coincidence events have been observed  during the 
effective observation time $T$. 
The probability 
of observing $n_c$ events, given a Poisson process 
of intensity
$r_b$, is 
\begin{equation}
P(n_c\,|\,r_b) = \frac{e^{-r_b\,T}\,(r_b\,T)^{n_c}}
                                 {n_c!}\,.
\end{equation}
Two remarks are now in order. First, one should 
be very careful about
calling $P(n_c\,|\,r_b)$ the `probability of the observed number 
of coincidences', because what has been observed is sure 
and no longer belongs to the domain of the uncertain, 
to which probability applies 
(the certain event has probability 1). 
$P(n_c\,|\,r_b)$ is, instead,
the probability of observing the 
hypothetical number of coincidences $n_c$, 
under the condition 
that the stochastic process is described by 
a Poisson of constant and 
precisely known intensity $r_b$ during the observation time $T$. 
Second, $P(n_c\,|\,r_b)$ does not provide, by itself, 
a result concerning what the researchers are interested in, 
i.e. the rate of g.w. bursts. 
In fact, $P(n_c\,|\,r_b)$ is a probabilistic statement about the 
possible outcome $n_c$, and not about the uncertain rate 
of g.w. bursts. 
However, it is 
rather intuitive that, if the observed number of coincidences
is of the order  
of the expected value of the background, i.e. 
$n_c \approx r_b\,T\pm\sqrt{r_b\,T}$, the background 
is considered to describe the outcome of the experiment well, 
while if $(n_c - r_b\,T)/\sqrt{r_b\,T}\gg 1$, 
suspicion is raised that some of the observed coincidences could be 
attributed to g.w. bursts
(or, more precisely,
 to any other physical effect not considered as background).
In this 
paper we consider that the only hypothesized but known 
background is a contribution of g.w. bursts.
As a consequence, there will 
be g.w. burst rates in which we  believe more, 
others in which we believe less, and others that we 
 rule out. 
In other words, we are faced with an inferential 
problem, which  must treated with care to avoid 
reporting the result in a way which might be misleading
(see e.g. examples given in Ref.~\cite{maxent98}). 

To give a numerical example, let us take the case 
of $\lambda_b=r_b\cdot T=100$, and let us assume that 130 
coincidences have been observed. 
The probability of $n_c=130$, 
given $\lambda_b=100$, is $6\cdot 10^{-4}$, but 
one should not say that `there is a probability of  $6\times 10^{-4}$
that the data come from the background'. 
In fact this would imply
that, `with 99.94\,\% probability, the data do not come from 
background', i.e. `they have to be attributed almost certainly  
to a genuine signal'. Indeed, the observation of a 
low-probability event 
does not imply that the hypothesis considered to be the  cause of it 
(the so-called `null hypothesis' $H_\circ$,
in our case  $H_\circ$ = `background is the only source of 
candidate events') has to be ruled out. 

One can recognize, behind the logic of standard hypothesis tests  
with which we are all familiar, a revised
version of the
classical proof by contradiction.
In standard dialectics,
one assumes a hypothesis
to be true, then looks for a logical consequence
which is manifestly false, in order to reject the hypothesis.
The `slight' difference introduced in the 
`classical' statistical tests
is that the false consequence is replaced by an
improbable one.
The argument might seem convincing at first sight, 
but it has no logical grounds. 
In fact, no matter how small the probability is, whatever
is observed is not in contradiction with the null hypothesis,
unless it is really impossible.  
This becomes self-evident when the 
probability of whatever can be observed is so small
that this kind of reasoning would rule out $H_\circ$ whatever 
one observes. For example, in our numerical example even 
$P(n_c=100\,|\,\lambda_b=100) = 4\,\%$ is below the 
standard probability level under which an event is declared 
`improbable'. This is  the reason 
why statisticians have invented 
`$p$-values', i.e. `probability
of the tail(s)' (see e.g. Ref.~\cite{pvalues}). 
For example, one would say, 
in our case, that the reason why the data are against the null hypothesis
is not simply because $P(n_c=130\,|\,\lambda_b=100)= 6\cdot 10^{-4}$, 
but because $P(n_c\ge 130\,|\,\lambda_b=100) = 0.23\,\%$. 
But this does not solve the problem, it makes it worse because 
one is considering  the conditional 
probability of not only what has actually been observed, but also 
what has not been observed (see e.g. 
Refs.~\cite{dagocern,BB}). 

Although we cannot presume to have been fully convincing with these very
brief critical remarks and therefore refer the reader to 
more general discussions on the subject (see e.g. Ref.~\cite{dagocern}
and references therein), the message is that one
is not allowed to  evaluate the probability of an effect
(or, even worse, the probability of an effect 
plus that of all rarer effects not actually observed), 
given a certain cause, and then to consider it 
as if it were the probability of the cause itself. 

Some readers might wonder why this paper is 
making such a big deal about
the criticism expressed above, which after all seems 
to be founded on intuition, logic and  good sense. 
The reason is that 
the standard education of physicists on the subject 
of probability  is based 
on a very peculiar and unnatural point of view  
(frequentism) which prevents probability of causes, 
i.e. what Poincar\'e 
calls {\it `the essential problem
of the experimental method'}~\cite{Poincare'}, 
 being talked about.
However, despite their education, 
physicists constantly make use of this concept, 
most of the time  correctly, as happens  
in  simple routine applications. 
But sometimes the combination of good intuition and 
unsuitable statistical approach yields wrong conclusions, 
as reported, e.g., in Ref.~\cite{maxent98}. 
Given this situation, we think that there is a strong probability that 
what we are going to say about the way of reporting results 
will be misunderstood, if it is not clear what is meant by probability
of causes (or of hypotheses, or of true values)
and how this can be evaluated on the basis of 
all available knowledge.  
Therefore, for the convenience of the reader, in the next section we give 
a short introduction to the problem of 
inference, extracted from Ref.~\cite{dagocern} and adapted 
to this context. 

\section{From data to true values}
\subsection{Learning from  observations: 
creating or modifying knowledge?}
Every measurement is made with the purpose of 
increasing  the knowledge
of the person who performs it, and of anybody else 
who may be interested in it, like
the members of a  scientific
 community.
It is clear that the need to perform a measurement indicates 
that one is in a state of uncertainty with respect to something, 
e.g. the value of a well-defined physics quantity. 
In all cases, the measurement has the purpose of modifying 
a given state of knowledge.
One might be tempted to say `acquire', instead of `modify', 
the state of knowledge, thus indicating that the knowledge could 
be created from nothing by the act of  measuring. 
However, it is not difficult to see that, in all cases, what 
we are dealing with is just an updating process, in the light 
of new facts and of some reason. 
To give an example about which 
everyone has good intuition, let us take the case of 
the measurement of the temperature in a room, 
using a digital thermometer (just to avoid uncertainty in the reading), 
and let us suppose that we get 21.7\,$^\circ$C. 
Although we may be uncertain about the tenths of a degree, 
there is no doubt that the measurement will have narrowed the 
interval of temperatures considered  possible
before the measurement: those compatible with the 
physiological feeling of a comfortable environment. 
According to our knowledge of the thermometer used,
or of thermometers in general, there will be values of temperature in a 
given interval around 21.7\,$^\circ$C in which we believe more and 
values outside the interval in which we believe less.
It is, however, also clear that if the thermometer had indicated, 
for the same physiological feeling, 17.3\,$^\circ$C, we might suspect that 
it was not well calibrated, while if it had indicated 2.5\,$^\circ$C 
we would have no doubt that the instrument was not working properly.  

The three cases correspond to three different degrees of 
modification of the knowledge. In particular, in the last case
the modification is null (but even in this case
we have learned something: the thermometer does not work!).

So, what makes us improve our knowledge after 
an empirical observation, is not the observation 
alone, but the observation framed in prior knowledge
about measurand and measurement.  Trained physicists 
always have such prior knowledge and often use it unconsciously. 
Imagine someone 
who has no scientific or technical education at all, entering 
 a physics laboratory and reading a number on an 
instrument: His scientific knowledge will not improve at all, apart 
from the triviality  that a given instrument  
displayed a number (not much of a contribution to knowledge!)

\subsection{From the probability of the 
observables to the pro\-ba\-bi\-li\-ty
of the true values}
Summarizing the argument so far, after having performed an experiment, 
which has resulted in the observed 
value $x$ being
 read on an instrument,
there are 
some values of the physical quantity (generically indicated by $\mu$)
 in which we believe more (we say `they
are more probable') and some others in which we believe less. 
The different possible 
true values can be characterized by a p.d.f. 
$f(\mu\,|\,x)$, conditioned by the observation $x$. 
To be more precise, $f(\mu)$ depends on many other pieces 
of information, like knowledge of the instruments, 
of the kind of measurement, and of reasonable values of $\mu$ 
to be expected. So using  K() to indicate the `knowledge', 
 to be precise, we should write 
$$f(\mu\,|\,x)\rightarrow
  f(\mu\,|\, x, \mbox{K(instr.)},\,\mbox{K(meas.)},\,\mbox{K}(\mu))\,,$$
although one often simply writes $f(\mu\,|\,x)$, or even $f(\mu)$, 
implicitly assuming the conditions. 

Clearly, $f(\mu\,|\,x)$ cannot be 
evaluated as relative frequencies of a long-run experiment. 
It  would be
absurd to imagine  a distribution of the values of  
$\mu$ for a given value $x$ read on the instrument, as
true values are not directly observable, being related to 
abstract concepts. 
Instead, 
relative frequencies can be used to evaluate the detector 
response:
$$f(x\,|\,\mu)\equiv 
f(x\,|\,\mu,\mbox{K(instr.),\,K(meas.)}).$$ 
This can be done either
by calibration with respect to a reference value 
or by Monte Carlo  simulation, 
as is currently done when $\mu$ refers
to a quantity for which a calibration cannot be 
made.\footnote{Note that a Monte Carlo 
program is nothing but a summary of the most reliable beliefs
about the phenomenology and the measurement process under study.}
More often, $f(x\,|\,\mu)$ is evaluated by reasonable assumptions, 
like when we assume a Gaussian model of error distribution,
or that the observed number of accidental coincidences
is described by a Poisson distribution. For example, 
one has to remember that, no matter how much  the off-timing 
distribution might be  Poisson-like, 
this empirical observation cannot be considered, 
strictly speaking, a proof,
having the same strength as mathematical theorem.
 Nevertheless, the fact that this observation 
will lead practically all researchers to believe a certain 
hypothesis makes it a typical example of the type of inferential process 
we are talking about.

The function $f(x\,|\,\mu)$ is usually called likelihood, since 
it quantifies how likely it is that $\mu$ will produce any given $x$.
Note that this function can be easily misinterpreted: As a probability
density function it is a function of $x$, since it describes the beliefs 
on $x$ for a given value of $\mu$. As a mathematical function it 
is also a function of $\mu$ in the sense that the p.d.f.
$f(x\,|\,\mu)$ depends on the `parameter' $\mu$. 
However, it is not correct to say that $f(x\,|\,\mu)$  
measures the belief that $x$  comes from $\mu$ 
 (in the sense that the observable $x$ has to be attributed to the 
true value $\mu$).  Instead, this degree of belief is denoted by
 $f(\mu\,|\,x)$.  
The confusion\footnote{Anticipating what will become 
clear in a while: if the prior is uniform, then 
$f(x\,|\,\mu)$ and $f(\mu\,|\,x)$ have identical mathematical expression.
This is the reason why the likelihood curve is often 
taken as if it were probability information about $\mu$.}    
 between  $f(\mu\,|\,x)$ and  $f(x\,|\,\mu)$
is a source of really terrible mistakes~\cite{maxent98,dagocern}.
 
So, the problem is how to get from the 
observation $x$ to $f(\mu)$. Before going to the formal
derivation of the formula to update
the beliefs, let us  try to justify intuitively the general rule, 
considering what happens when $\mu$ can assume only two values. 
If they seem to us equally possible, 
it is natural to favour the value which gives 
the highest likelihood of producing $x$. For example, assuming 
$\mu_1=-1$, $\mu_2=10$, 
considering
a Gaussian likelihood with $\sigma = 3$, and having observed $x=2$, 
one would tend to believe that the observation is most likely
caused by $\mu_1$.
If, on the
other hand, we add the extra information 
that the quantity of interest is positive, 
then $\mu_1$ is no longer the most probable cause 
but an impossible one;
$\mu_2$ becomes certain. 
There are, in general, intermediate cases in which, 
because of  previous knowledge, 
one tends to believe {\it a priori} 
more in one or other of the causes.
(For example, one could imagine a small Monte Carlo in which 
$\mu_1$ and $\mu_2$ are randomly chosen with probability ratio
1 to $10^5$: where, then, does $x=2$ come from?)
It follows that,
in the light of a new observation, the degree of belief of a given 
value of $\mu$ will be proportional to 
\begin {itemize}
\item
the likelihood that $\mu$ will 
produce the observed effect; 
\item 
the degree of belief attributed
to $\mu$ before the observation, quantified by $f_\circ(\mu)$ (the 
 `prior').  
\end{itemize}
We have finally
\begin{equation}
f(\mu\,|\,x)\propto f(x\,|\,\mu)\cdot f_\circ(\mu)\,.
\label{eq:Bayes}
\end{equation}
This is one of the ways of writing  Bayes' 
theorem. The proportionality factor is simply given by the normalization
of $f(\mu\,|\,x)$ to 1. 
At this point it is important 
to write Bayes' formula once more, making all conditions explicit: 
$$f(\mu\,|\,x\,,\mbox{K(instr.),\,K(meas.),\,K}_\circ(\mu)) 
  \propto f(x\,|\,\mu,\mbox{K(instr.),\,K(meas.)})\cdot 
  f(\mu\,|\,\mbox{K}_\circ(\mu))\,,$$ 
where $f(\mu\,|\,\mbox{K}_\circ(\mu))\equiv f_\circ(\mu)$. 

\subsection{Derivation of Bayes' theorem from a physicist's perspective}
The concepts illustrated 
in the previous section can be formalized 
using the following reasoning. 
\begin{itemize}
\item 
Before doing the experiment we are uncertain on $\mu$ and on 
$x$: we know neither the true value, nor
the observed value. Generally speaking, this uncertainty is 
quantified by $f(\mu, x)$.
\item
Under the hypothesis that we observe $x$, we can calculate
the conditional probability 
\begin{equation}
f(\mu\,|\, x) 
= \frac{f(\mu, x)}
       {f(x)}
= \frac{f(\mu, x)}
       {\int \!f(\mu, x)\,\mbox{d}\mu}\,.
\end{equation}
\item
At this point, it seems we are stuck, because we are usually
more uncertain about $\{\mu, x\}$ than about $\mu$. 
However, we note that  $f(\mu, x)$ 
can be calculated from $f(x\,|\,\mu)$ 
and $f(\mu)$: 
\begin{equation}
f(\mu, x) = f(x\,|\,\mu)\cdot f(\mu)\,.
\end{equation}
This is the key observation to solve our problem.
\item 
If we do an experiment we need to have a good idea of the 
behaviour of the apparatus, therefore $f(x\,|\,\mu)$ 
must be a narrow distribution. The most uncertain 
contribution remains the prior knowledge about $\mu$, quantified by 
$f_\circ(\mu)$ (the subscript is to remind us that this is 
a prior  about $\mu$). Note that 
it is all right that $f_\circ(\mu)$ is rather broad (`vague'), 
because we want to learn about $\mu$ itself, performing an experiment
with an apparatus having a narrow response around true values. 
\item
Putting all the pieces together we get the standard formula
of  Bayes' theorem for 
uncertain quantities:
\begin{equation}
f(\mu\,|\,x) = 
\frac{f(x\,|\,\mu)\cdot f_\circ(\mu)}
     {\int \! f(x\,|\,\mu)\cdot f_\circ(\mu)\, \mbox{d}\mu}\,.
\end{equation}
\end{itemize}
The steps followed in this proof of the theorem should 
convince the reader that $ f(\mu\,|\,x)$ calculated 
in this way is the most we can say about $\mu$
with the given state of information. 

One may be worried about the presence of $f_\circ(\mu)$ in the 
result; but this is simply unavoidable, and for this reason 
we should be relaxed about it~\cite{priors}.  
$f_\circ(\mu)$ becomes irrelevant in routine cases
because the likelihood is usually 
very narrow 
(when seen as a mathematical function of $\mu$) 
with respect to $f_\circ(\mu)$, such 
that the prior is reabsorbed in the normalization 
factor.\footnote{To make it clear, 
if one wants to measure the temperature in a  room one does not
choose a thermometer with an r.m.s. error of 5 degrees, 
because one's physiological prior is more accurate than what 
could be learned from such a  measurement;  
but the same instrument provides useful information at 200$^\circ$C
or at -50$^\circ$C.} In contrast, 
in the sophisticated frontier-science measurements $f_\circ(\mu)$ 
does matter, as will be illustrated below.

Let us  conclude this very short introduction on Bayesian inference.
\begin{itemize}
\item
It is impossible to give a probabilistic result on a physics quantity 
without passing through priors (``it is impossible 
to make the inferential omelette without 
breaking the Bayesian egg'',  some like to say).
\item
When the probabilistic result seems not to depend on priors, 
we are in a condition (or we make the tacit assumption!)
in which the prior distribution acts as a constant; 
this approximation
is very good when the likelihood is much narrower than
the prior, as usually happens in routine measurements.
\item
In frontier science, priors must be considered with much care,
and this will be the main task of this paper.
\end{itemize}

\section{Inferring gravitational-wave burst rate}\label{inferring}
\subsection{Modelling the inferential process}
Now that the inferential scheme has been set up, let us
rephrase our problem in the language of Bayesian statistics:
\begin{itemize}
\item
the physical quantity of interest, and with respect to which 
we are in a state of great uncertainty, is the  
g.w. burst rate $r$;
\item
we are practically sure about the expected rate of background 
events $r_b$ (but not about the number which will  
actually be observed);
\item
what is certain is the number $n_c$ 
of coincidences which have been observed 
(stating that the observed
number of coincidences is $n_c\pm\sqrt{n_c}$ does not make any sense!), 
although we do not
know how many of these events have to be
attributed to background and how many (if any) to g.w. bursts.
\end{itemize}
For a given hypothesis $r$  
the number of coincidence events which can 
be observed in the observation time $T$ is described by a Poisson 
process having an intensity which is the sum
of that due to background and that due to signal. Therefore the likelihood
is
\begin{equation}
f(n_c\,|\,r,r_b) =\frac{e^{-(r+r_b)\,T}((r+r_b)\,T)^{n_c}}{n_c!}\,,
\label{eq:likr}
\end{equation}
and, making use of  Bayes' theorem, we get
\begin{equation}
f(r\,|\,n_c,r_b) \propto 
\frac{e^{-(r+r_b)\,T}((r+r_b)\,T)^{n_c}}{n_c!} f_\circ(r)\,.
\label{eq:frfin}
\end{equation}

At this point we are faced with the problem of what 
$f_\circ(r)$ to choose. The best way of understanding why
this choice can be troublesome is to illustrate the problem 
with numerical examples. 
Let us consider $T$ as unit time (e.g. one month), 
a background rate $r_b$ such that $r_b\times T=1$,
and the following hypothetical observations: 
 $n_c=0$;  $n_c=1$; $n_c=5$.

\subsection{Uniform prior} 
\begin{figure}[t]
\begin{center}
\epsfig{file=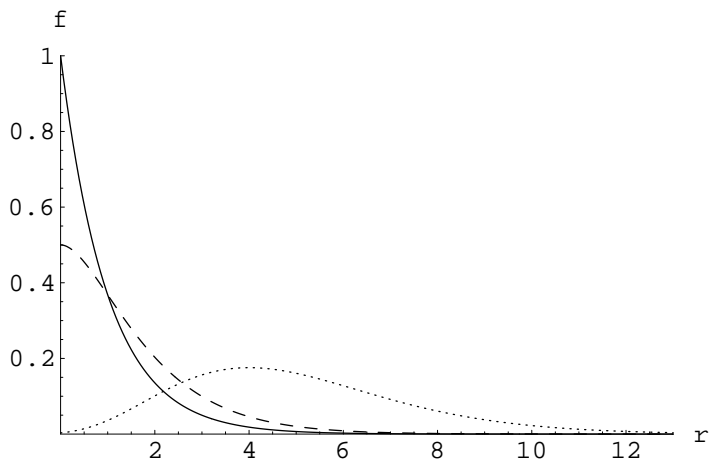,clip=,width=0.6\linewidth} 
\end{center}
\caption{\small Distribution of the values of the rate 
$r$, in units of events/month, inferred from an expected 
rate of background events $r_b=1$ event/month, 
an initial uniform distribution $f_\circ(r)=k$
and the following numbers of observed events: 
 0 (continuous);  1 (dashed);  5 (dotted).}
 \label{fig:bayesflat}
\end{figure}

One might think that a good 
`democratic'
choice would be a uniform distribution in $r$, 
i.e. $f_\circ(r)=k$. Inserting this prior in (\ref{eq:frfin})
and normalizing the final distribution we get
(see e.g. Ref.~\cite{dagocern})
\begin{equation}
f(r\,|\,n_c,r_b,f_\circ(r)=k) =
\frac{e^{-r\,T}((r+r_b)\,T)^{n_c}
     }
     {n_c!\,\sum_{n=0}^{n_c}\frac{(r_b\,T)^n}{n!}
     }\,.
\end{equation}
The resulting final distributions are shown in 
Fig.~\ref{fig:bayesflat}.
For $n_c=0$ and 1 the distributions are peaked at zero, while
for $n_c=5$ the distribution appears so neatly separated from 
$r=0$ that it seems a convincing proof that the 
postulated physics process searched-for does exist. 
In  the cases $n_c=0$ and 1, researchers usually  
present the result with an upper limit (typically 95\,\%), 
on the basis that $f(r)$ seems compatible with no effect, 
as suggested by Fig.~\ref{fig:bayesflat}.
For example, in the simplest and well-known  case 
of $n_c=0$ the 95\,\% C.L. upper limit
is 3 events/month. The usual meaning one attributes
to the limit is that, if the physics process of interest
exists, 
then there is a  95\,\% probability  
that its rate is below  3 events/month,
resulting from
\begin{equation}
\int_0^3\!\!f(r\,|\,n_c=0,r_b=1,f_\circ(r)=k)\,\mbox{d}r = 0.95\,.
\label{eq:limit}
\end{equation}
But there are other infinite probabilistic statements that can be
derived from $f(r\,|\,r_b,n_c=0)$. For example, 
$P(r>3\,\mbox{events/month}) = 5\,\%$, 
$P(r>0.1\,\mbox{events/month})=90\,\%$, 
$P(r>0.01\,\mbox{events/month})=99\,\%$, and so on. 
Without doubt, researchers will not hesitate to publish 
the 95\,\% upper limit, but they would feel uncomfortable stating
that they believe 99\,\% that, if the g.w. bursts exist at all, then 
the rate is above 0.01 events/month.\footnote{If one 
assesses a probability value of 99\,\%, one
should be as confident that the event will turn out to be true
as one would be of extracting a white 
ball from an urn containing 99 white balls 
and one black. If this is not the case, one is, 
consciously or not, responsible for misinformation.
So, other people, 
trusting the person who made that probability assessment, 
will form their opinion and make their decisions using a 
probability value which does not correspond to what that
person believes.}
The reason for this uneasiness can be found 
in the uniform prior, which might 
not correspond to the prior knowledge that 
researchers really have. 
Let us, then, examine more closely the meaning of the uniform
distribution and its consequences. Saying that 
$f_\circ(r)=k$, means that 
$\mbox{d}P/\mbox{d}r = k$, i.e. $P\propto \Delta r$;
for example,
\begin{equation}
P(0.1\le r\le 1) = \frac{1}{10}\,P(1\le r\le 10) =
\frac{1}{100}\,P(10\le r\le 100)\ldots\,,  
\label{eq:uniforme}
\end{equation}
and so on. But, taken literally, 
this prior is hardly ever  reasonable. The problem
is not due to  the divergence for $r\rightarrow \infty$
 which makes $f_\circ(r)$ not normalizable (these kinds of distributions
are called `improper'). This 
mathematical nuisance is automatically cured when 
$f_\circ(r)$ is multiplied by the likelihood, which, for a finite
number of observed events, 
vanishes rapidly enough for  $r\rightarrow \infty$. 
A much more serious  problem is related to the fact
that the uniform distribution assigns to
all the infinite orders of 
magnitude left of 1 a probability which is only 
1/9  of the probability of the decade between 
1 and 10, or 1\,\% of the probability of the first two decades,
and so on.
This is the reason why, even if no coincidence 
events have been observed, the final distribution 
obtained from zero events observed
(continuous curve of Fig.~\ref{fig:bayesflat})
implies that $P(r\ge 1\,\mbox{event/month}) = 37\,\%$. 

\subsection{Jeffreys' prior}
A prior distribution alternative to the uniform can be based
on the observation that what often seems uniform is not the probability
per unit of $r$, but rather the probability per decade of $r$, 
i.e. researchers may feel equally uncertain about 
the orders of magnitudes of $r$, namely
 \begin{equation}
P(0.1\le r\le 1) = P(1\le r\le 10) =
P(10\le r\le 100)\ldots\,. 
\label{eq:Jeffreys}
\end{equation}
This implies that $\mbox{d}P/\mbox{d}\ln r = k$, 
or $\mbox{d}P/\mbox{d}r\propto 1/r$. 
This prior is known as  Jeffreys' prior~\cite{Jeffreys}, 
and it is indeed very interesting, at least from a very 
abstract point of view (though it tends to be 
misused, as is discussed in Ref.~\cite{priors}). 
If we take Jeffreys' prior literally, 
it does not work in our case either. In fact, when inserted 
in Ref.~(\ref{eq:frfin}), it produces a divergence for $r\rightarrow 0$. 
This  is due to the infinite orders of 
magnitude left of 1, to each of which we give equal prior 
probability, and to the fact that the likelihood 
(\ref{eq:likr}) goes to
a constant for $r\rightarrow 0$. Therefore, for any $r_\circ>0$, 
we have $P(r<r_\circ)/P(r>r_\circ)=\infty$. 
To get a finite result we need a cut-off at a given $r_{min}$.
\begin{figure}
\begin{center}
\begin{tabular}{cc}
\epsfig{file=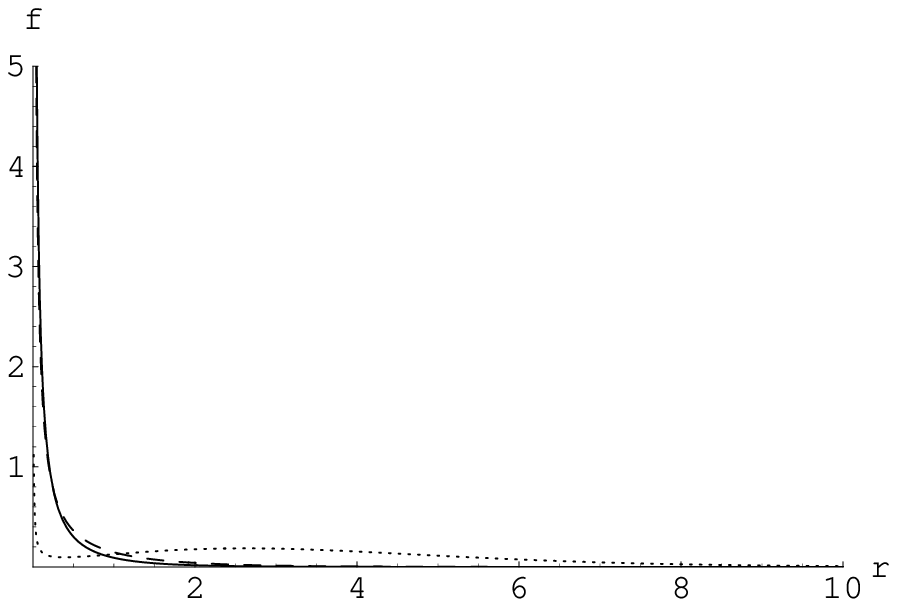,width=0.48\linewidth,height=6.6cm,clip=} &
\epsfig{file=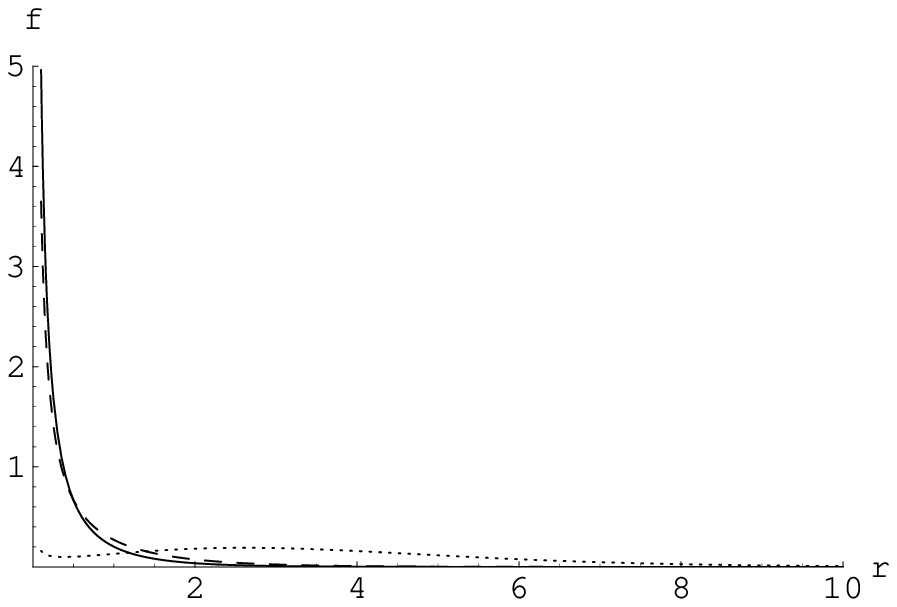,width=0.48\linewidth,height=6.6cm,clip=} 
\end{tabular}
\end{center}
\caption{\small  Final distributions for the same 
experimental configuration
of Fig.~\ref{fig:bayesflat}, but with a Jeffreys' prior 
with cut-off at $r_{min}=0.01$\,events/month (left plot)
and  $r_{min}=0.1$\,events/month (right plot).}
\label{fig:Jeffreys}
\end{figure}

As an exercise, just to get a feeling of both
the difference with respect 
to the case of the uniform distribution, and the dependence 
on the cut-off, we report in Fig.~\ref{fig:Jeffreys}
the results obtained for the same experimental conditions
as Fig.~\ref{fig:bayesflat}, but with a Jeffreys' prior 
truncated at $r_{min} =0.1$ and 0.01. One can see that the 
final distributions conditioned by 0 or 1 events observed 
are pulled towards $r=0$ by the new priors, while the case 
of $n_c=5$ is more robust, although it is no longer 
nicely separated from zero.

\subsection{Role of priors}
The strong dependence of the final distributions on the priors 
shown in this example 
should not be considered a bad feature, 
as were an artifact of Bayesian 
inference. Putting it the other way round, the Bayesian inference 
reproduces, in a formal way, what researchers 
already have clear in their minds 
as a result of intuition and experience. In the numerical examples
we are dealing with, the dependence of the final distributions 
on the priors 
is just a hint of the fact that
the experimental data are not so strong as to lead 
every scientist to the same conclusion
(in other words, the experimental and theoretical situation is far 
from the well-established 
one upon which intersubjectivity is based). 
For this reason, one should worry, instead, about statistical
methods which advertise `objective' 
probabilistic results in such a critical situation.  

\begin{figure}[!b]
\begin{center}
\epsfig{file=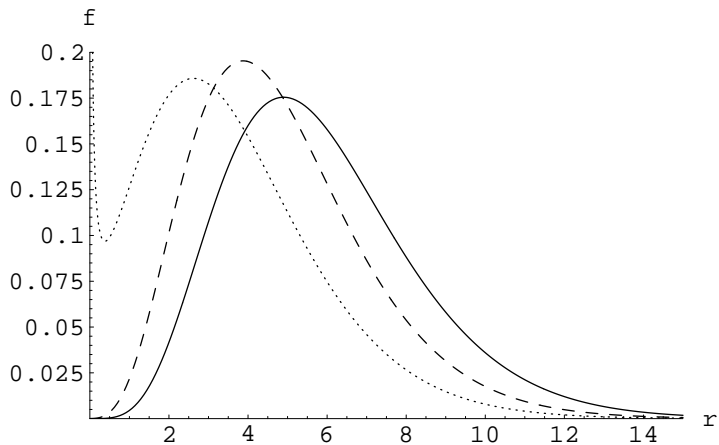,clip=,width=0.6\linewidth} 
\end{center}
\caption{\small Distribution of the values of the rate 
$r$, in units of events/month, inferred from 
five observed events, an expected 
rate of background events $r_b=0.1$ events/month, 
and the following priors: 
uniform distribution $f_\circ(r)=k$ (continuous); 
Jeffreys' prior truncated at $r_{min}=0.01$ (dashed). 
The case of the Jeffreys' priors is also reported for $r_b=1$
event/month (dotted).} 
 \label{fig:lowbgd}
\end{figure}

When the experimental situation is more solid, as for example
in the case of five events observed out 
of only 0.1 expected from background,
the conclusions become very similar, virtually independent 
of the priors (see Fig.~\ref{fig:lowbgd}), 
unless the priors reflected really widely differing opinions.

The possibility that scientists might have distant and almost 
non-overlapping  priors, such that agreement is reached only after 
a huge amount of very convincing data, 
 should not be overlooked, as this is, in fact, the typical
situation in frontier research.
Even 100 events observed
out of 0.1 expected from background are not a 
logical proof of the existence 
of bursts, since the observation is not in contradiction 
with the background alone. Nevertheless, any reasonable physicist
will agree that this is highly unlikely
(we shall come back to evolution of beliefs in 
Section \ref{ss:evolution}.) 

\subsection{Priors reflecting the positive attitude of 
researchers}\label{ss:pa}
Having clarified the role of priors 
in the assessment of probabilistic statements about 
true values, and their critical influence on 
frontier-research results, it is clear that, in our opinion, 
``reference priors do not exist''~\cite{priors,Bernardo}.
However, we find   
that the ``concept of a `minimal informative' prior specification
- appropriately defined!''\,\cite{BS} can sometimes be useful, 
if the practitioner 
is aware of the assumptions behind the specification. 

We can now ask ourselves what would be a prior specification 
common to rational and responsible people
 who have planned, financed and operated
frontier-type experiments. This is what we call the 
positive attitude of researchers~\cite{dagocern}. 
Certainly, 
the researchers believed there was a good chance, depending
on the kind of measurement, that they would end up with a number of candidate 
events well above the background; or that the  physical quantity 
of interest was well above the experimental resolution; 
or that a certain rate would be in the region of sensitivity. 
One can show that the results obtained with reasonable 
prior distributions, chosen to model this positive attitude, 
are very similar to those obtainable by an improper uniform prior
and, in particular, the upper/lower bounds obtained are 
very stable (see Sections 5.4.3 and 9.1.1 of Ref.~\cite{dagocern}). 

\begin{figure}[!h]
\begin{center}
\begin{tabular}{|c|}\hline
\epsfig{file=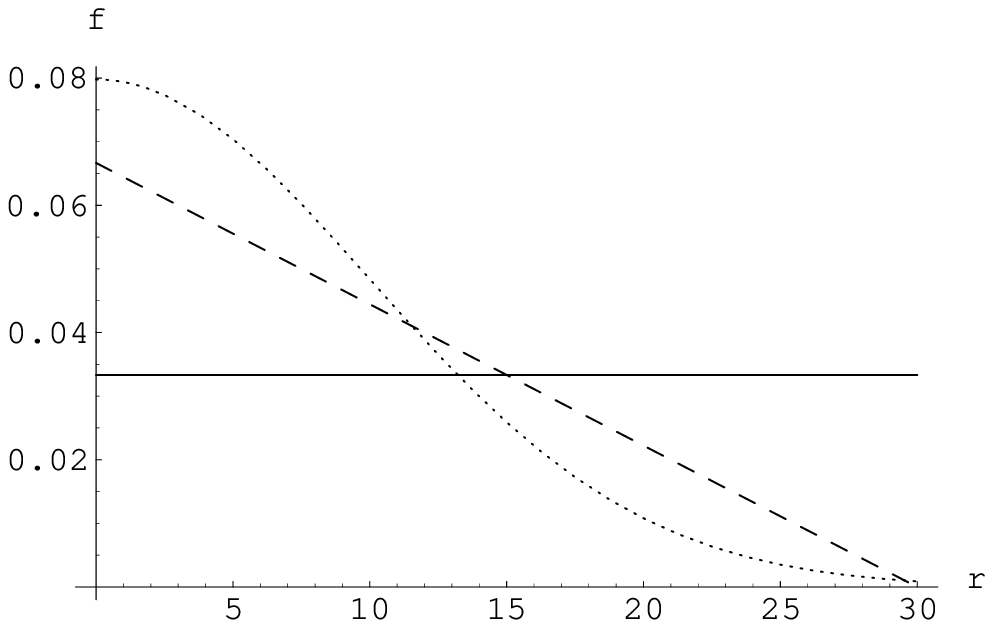,width=0.65\linewidth,clip=} \\ \hline
\epsfig{file=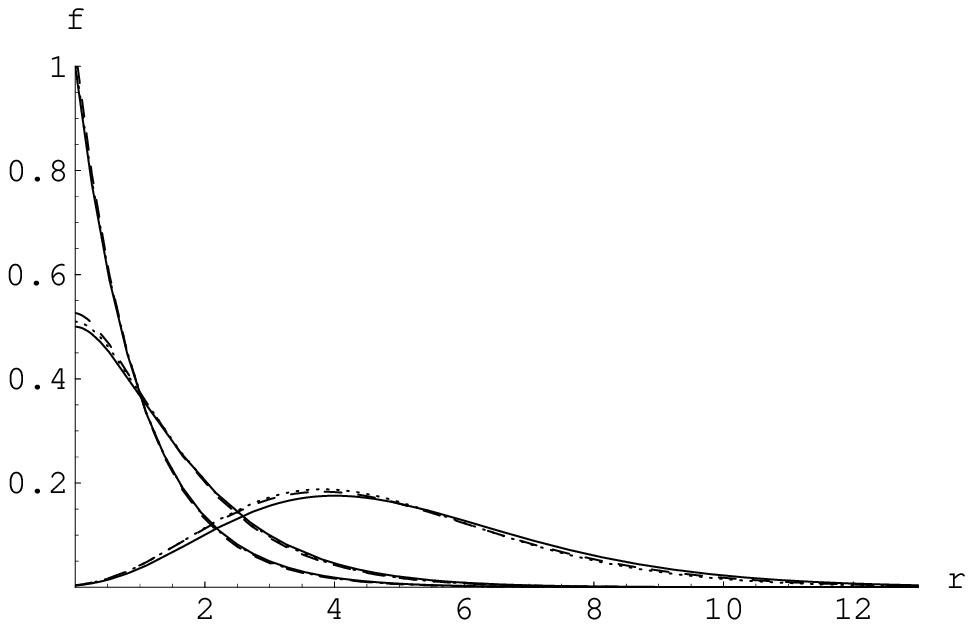,width=0.65\linewidth,clip=} \\ \hline
\end{tabular}
\end{center}
\caption{\small The upper plot shows some 
reasonable priors reflecting the positive 
attitude of researchers: uniform distribution (continuous);
triangular distribution (dashed); half-Gaussian distribution (dotted). 
The lower plot shows how the results of 
Fig.~\ref{fig:bayesflat}, obtained
starting from an improper uniform distribution,
(do not) change if, instead,  the
priors of the upper plot are used.} 
\label{fig:pa}
\end{figure}

Let us apply this idea to the exercise we are dealing with: 
0, 1 or 5 events observed over a background of 1 event
(Fig.~\ref{fig:bayesflat}). Searching for a rare process 
with a detector having 
a background of 1 event/month, for an exposure time of 
one month, a positive attitude would be to think  
that signal rates of several events per month are
rather possible. On the other hand, the fact that the process is
considered to be rare implies that one does not expect a 
very large rate (i.e. large rates would contradict 
previous experimental information), 
and also that there is some belief that
 the rate could be very small, 
virtually zero. Let us assume that the researchers are almost sure that
the rate is  below 30 events/month. 
 We can consider, as examples, the following prior distributions.
\begin{itemize}
\item
A uniform distribution between 0 and 30: 
    \begin{equation}
    \mbox{ }\hspace{0.8cm} f_\circ(r) = 1/30 
    \hspace{3.3cm}(0\le r\le 30).
    \end{equation}
\item
A triangular distribution: 
     \begin{equation}
    \mbox{ }\hspace{0.9cm}f_\circ(r) = \frac{1}{450}\,(30-r)
    \hspace{1.9cm}(0\le r\le 30).
    \end{equation}
\item
A half-Gaussian distribution of $\sigma_\circ=10$
       \begin{equation}
    f_\circ(r) = \frac{2}{\sqrt{2\pi}\,\sigma_\circ}
    \exp\left[-\frac{r^2}{2\,\sigma_\circ^2}\right] 
    \hspace{0.5cm}(r \ge 0).
    \end{equation}
\end{itemize}
The last two functions model the fact that researchers might believe
that small values of $r$ are more possible than high values,
as is often the case.  Moreover, 
the half-Gaussian distribution also models the more realistic
belief that rates above 30 events/month are not excluded, 
although they are considered very unlikely.\footnote{
We will see in Section \ref{ss:turning} that realistic 
priors can be roughly modelled by a log-normal distribution. 
With parameters chosen to describe 
the positive attitude we are considering, this distribution
would give results 
practically equivalent to the three priors we are using now.}  
The three priors are shown in the upper plot of Fig.~\ref{fig:pa}. 
The resulting final distributions are shown in the lower plot of the
same figure. The three solutions are practically indistinguishable,
and, in particular, very similar to the results obtained 
by an improper uniform distribution (Fig.~\ref{fig:bayesflat}).
This suggests that the 
improper uniform prior represents a practical and easy
way of representing the prior specification for this kind 
of problem if one assumes what we have called the positive attitude of 
the researchers. Therefore, this prior could represent a way of 
reporting conventional probabilistic results, if one is 
aware of the limits of the convention. Seeking a truly 
objective probabilistic result --- we like to stress the concept 
again --- 
is a dream.
\vspace{2.0cm}

\newpage
\section{Prior-free presentation of the experimental evidence}
\label{sec:bur}
At this point, we want to reassure 
the reader (who we imagine at this point to be swirling around
in the stormy sea of subjectivism) that it is possible 
to present data in an `objective' way, 
on the condition that all thoughts of  providing 
probabilistic results about the measurand are suspended.

Let us take again Bayes' theorem (Eq.\,(\ref{eq:Bayes})),
which we rewrite here in terms of the uncertain quantities of
interest
\begin{equation}
f(r\,|\,n_c,r_b) \propto f(n_c\,|\,r,r_b)\cdot f_\circ(r)\,,
\label{eq:Bayes1}
\end{equation}
and consider only two possible values of $r$, let them be $r_1$ and $r_2$.
From (\ref{eq:Bayes1}) it follows that
\begin{equation}
\frac{f(r_1\,|\,n_c,r_b)}{f(r_2\,|\,n_c,r_b)} = 
\underbrace{\frac{f(n_c\,|\,r_1,r_b)}
                 {f(n_c\,|\,r_2,r_b)}
           }_{\mbox{\it Bayes factor}}\,\cdot\, 
\frac{f_\circ(r_1)}{f_\circ(r_2)}\,.
\label{eq:Bayesf}
\end{equation}
This is a common way of rewriting  the result of the Bayesian 
inference for a couple of hypotheses, keeping 
the contributions due to the experimental evidence
and to the prior knowledge separate. The ratio 
of likelihoods is known as the Bayes factor and it quantifies 
the ratio of evidence provided by the data in favour 
of either hypothesis. The  Bayes factor is considered 
to be practically objective because  
likelihoods (i.e. probabilistic description of 
the detector response) are usually 
much less critical than priors about the physics quantity of 
interest.\footnote{Note that this assumption might be questionable
in the  sophisticated field of g.w. search. For example, the 
effects of local sources of noise in the detectors are 
not well understood. 
This is what makes a substantial difference between 
a single detector and a coincidence experiment. 
The likelihood function summarizes the best knowledge 
about the g.w. burst detection and identification, and about
noise behaviour, the tail of which can be very critical. 
In a coincidence experiment the detailed knowledge 
of the background becomes 
uncritical, as the only relevant hypothesis which makes
accidental coincidence described by a Poisson distribution 
is the stationarity of the experimental conditions over the 
considered observation time.}   

The Bayes factor can be extended to a continuous set of hypotheses $r$, 
considering a function which gives the Bayes factor 
of each value of $r$ with respect to a reference value $r_{REF}$. 
The reference value could be arbitrary, but for our problem  the choice 
$r_{REF}=0$, giving 
\begin{equation}
{\cal R}(r;n_c,r_b) = \frac{f(n_c\,|\,r,r_b)}{f(n_c\,|\,r=0,r_b)}\,,
\label{eq:rbur_def}
\end{equation}
 is very convenient for comparing and combining 
the experimental results~\cite{ci,zeus,higgs}. 
The function ${\cal R}$ has nice 
intuitive interpretations which can be highlighted by 
reordering the terms of (\ref{eq:Bayesf}) in the form
\begin{equation}
\frac{f(r\,|\,n_c,r_b)}{f_\circ(r)}\left/
\frac{f(r=0\,|\,n_c,r_b)}{f_\circ(r=0)}\right. = 
\frac{f(n_c\,|\,r,r_b)}{f(n_c\,|\,r=0,r_b)} = {\cal R}(r;n_c,r_b)
\label{eq:rbur}
\end{equation}
(valid for all  possible a priori $r$ values). 
${\cal R}$ has the probabilistic interpretation of relative belief
updating ratio, or the geometrical interpretation
of shape distortion function of the probability density function. 
 ${\cal R}$ goes to 1 for $r\rightarrow 0$, i.e. in the asymptotic region
in which the experimental sensitivity is lost: As long as it is 1,
the shape of the p.d.f. (and therefore the relative probabilities in that 
region) remains unchanged.  Instead,  in the limit 
${\cal R}\rightarrow 0$ (for large $r$) 
the final p.d.f. vanishes, i.e. the beliefs go 
to zero no matter how strong they were before. 
In the case of the Poisson process we are considering, the relative
belief updating factor becomes
\begin{equation}
{\cal R}(r;n_c,r_b,T) = e^{-r\,T}\left(1+\frac{r}{r_b}\right)^{n_c}\,,
\label{eq:erre}
\end{equation}
with the condition\footnote{The case $r_b=n_c=0$ 
yields ${\cal R}(r) = e^{-r}$, obtainable starting directly 
from Eq. (\ref{eq:rbur_def}), defining ${\cal R}$, and from 
Eq. (\ref{eq:likr}), giving the likelihood. Also the case 
$r_b\rightarrow \infty$ has to be evaluated directly from 
the definition of ${\cal R}$ and from the likelihood, 
yielding ${\cal R}=1\ \forall\, r$; finally, 
the case $r_b=0$
and $n_c>0$ makes $r=0$ impossible, thus prompting 
a claim for discovery --
and it no longer makes sense for the ${\cal R}$ function 
defined above to have that nice 
asymptotic behaviour in the insensitivity region.} $r_b>0$ if  $n_c >0$.  

\begin{figure}
\begin{center}
\epsfig{file=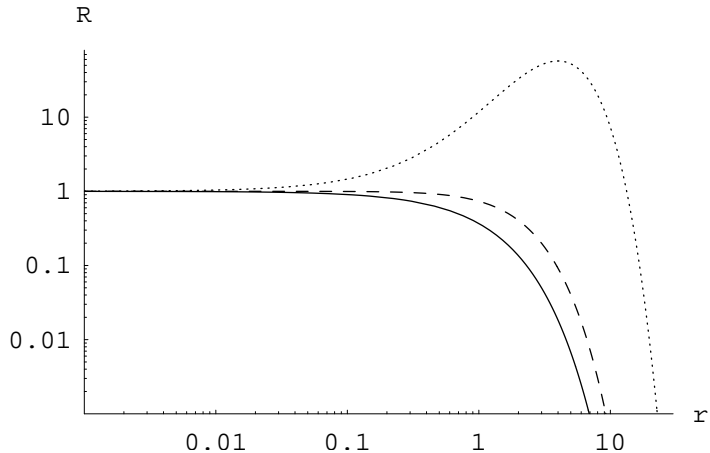,clip=,width=0.60\linewidth} 
\end{center}
\caption{\small Relative belief updating ratio ${\cal R}$
for the Poisson intensity parameter $r$ for the cases 
of Fig.~\ref{fig:bayesflat}.}
\label{fig:rbur}
\end{figure}

Figure \ref{fig:rbur} shows the ${\cal R}$ function for 
the numerical examples considered above. The abscissa 
has been drawn in a log scale to make it clear that several orders 
of magnitude are involved. 
These curves transmit the result of the experiment 
immediately and intuitively:
\begin{itemize}
\item
whatever one's beliefs on $r$ were before the data, these curves
show how one must\footnote{It  really is a `must' and not 
a `suggestion'. In fact, although probabilities may depend
on individuals (`subjective'), the way they are updated 
follows from standard logic (yielding Bayes' theorem) and thus 
is `objective'.} change them; 
\item
the beliefs one had 
for rates far above 20 events/month are killed by the 
experimental result;
\item
if one believed strongly that the rate had to be below 
0.1 events/month, the data are irrelevant;
\item
the case in which no candidate events
have been observed gives the strongest constraint on the 
rate $r$;
\item
the case of five candidate events over an expected background of
one produces a  
 peak of ${\cal R}$ which  corroborates the beliefs 
around 4 events/month only if there were sizable 
prior beliefs in that region.  
\end{itemize}
Moreover there are some technical advantages in reporting 
the ${\cal R}$ function as a result of a search experiment. 
\begin{itemize}
\item
One deals with numerical values which can differ from unity 
only by a few orders of magnitude in the region of interest, 
while the values of the likelihood can be extremely low.
For this  reason, the comparison between different 
results given by the ${\cal R}$ function can be perceived 
better than if these results were published in terms of likelihood.
\item
Since ${\cal R}$ differs from the likelihood only 
by a factor, it can be used directly in  Bayes' theorem, 
which does not depend on constants, whenever 
probabilistic considerations are needed.\footnote{Note that, although 
it is important to present prior-free results, at a 
certain moment a probability assessment about $r$ 
can be important,
for example, in forming one's own idea about the most likely range
of $r$, or in taking decisions
about planning and financing of future experiments.}
In fact, 
\begin{equation}
f(r\,|\,n_c\,r_b) \propto {\cal R}(r; n_c, r_b)\cdot f_\circ(r)\,.
\end{equation}
\item
The combination of different independent 
results on the same\footnote{See comments about  
the choice of the energy threshold in Section \ref{sec:commenti}.} 
quantity $r$ 
can be done straightforwardly by multiplying  individual
${\cal R}$ functions:
\begin{equation}
{\cal R}(r;\mbox{all data}) = \Pi_i {\cal R}(r;\mbox{data}_i)\,.
\label{eq:bur_comb}
\end{equation}
\item
Finally, one does not need to decide a priori if one wants to make a
`discovery' or an `upper limit' analysis as 
conventional statistics teaches (see e.g. criticisms in 
Ref.~\cite{BB}): 
the ${\cal R}$ function
represents the most unbiased way of presenting the results
and everyone can draw their own conclusions.
\end{itemize}

\section{A case study based on realistic detector performances}
\label{esempi}
\subsection{Prior-free results}\label{ss:priors-free}
We now give a numerical example which uses the realistic 
parameters of an actual g.w.
antenna and simulates possible experimental outcomes
that g.w. researchers could be faced with. 
One of the best performances, in terms of 
sensitivity and duty cycle, was
obtained with the Explorer antenna in 1991
\cite{bardonecchia,explorer}. The antenna worked with a duty cycle
of $67\%$ for $\simeq 180$ days, i.e. 
122 effective days, at a noise level 
of $\simeq 8$ mK (that is in terms of signal amplitude 
$h=7\cdot 10^{-19}$). 
At the chosen threshold, $h= 2.5\cdot 10^{-18}$, the event rate was
roughly 100 events/day. We shall take this as the 
reference value of the background rate for our numerical examples.

\begin{figure}
\centering\epsfig{file=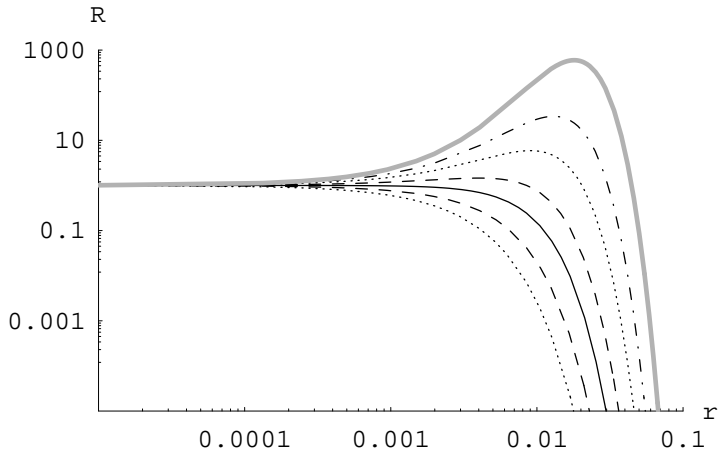,width=0.6\linewidth,clip=}
\caption{\small Belief updating ratios on a log-log
scale for different observations.
The abscissa shows the signal rate, in events per day. 
The continuous curve correspond to an observation
equal to the background, the other curves to 
a difference between observation and background 
expectation ranging from $-2$ to $+4$ standard deviations.
The grey curve (+4 st. dev.) is the case that will be studied 
in more detail (see Figs. \ref{fig:priors}, 
\ref{fig:finals} and \ref{fig:seq}).}
\label{fig:burloglog}
\end{figure}

We now imagine a coincidence analysis between two antennae
having the 1991 Explorer characteristics, parallel 
to each other, far enough apart not to be sensitive
to the same local effects and 
being operated  for 1000 days.
We consider here a fixed window of  0.2\,s
(see e.g. Refs. \cite{bardonecchia} and \cite{nostro2}),
yielding 
an expected number 
of accidental coincidences of
$r_b=100 \times 100 \times 0.2/86400=0.02$ events/day.
The expected number of accidental coincidences is, then, 
$\mbox{E}[n_c\,|\,r_b, T]=  r_b\,T = \lambda_b = 20$
(the product $r_b\, T$ will be indicated hereafter by $\lambda_b$).  
Let us consider the following 
numbers of observed coincidences: 
$n_c = 10$, 15, 20, 24, 29, 33, 38, roughly corresponding 
to a difference between observation and background 
expectation ranging
from $-2$ to $+4$ standard deviations 
[$\sigma(n_c\,|\,\lambda_b)=\sqrt{20}$]. 
The corresponding ${\cal R}$ values are shown  in 
Fig.~\ref{fig:burloglog}. 
We see that all results exclude rate values above
$\approx 0.1$ bursts/day, while the experiment loses 
sensitivity below $\approx 0.001$  bursts/day. 
In the case of excess of observed coincidences above the 
background expectation ($n_c>\lambda_b$), 
the ${\cal R}$ function has a peak 
at $r_m = n_c/T-r_b$, with a peak value of 
${\cal R}_m = e^{-r_m}(n_c/\lambda_b)^{n_c}$. 
The peak rises very rapidly with $n_c$. For example,
for $n_c=38$ the peak value is roughly 600, at
 $r_m=1.8\cdot 10^{-2}$ bursts/day. 

\subsection{Turning the results into probabilities}\label{ss:turning}
A peak value of 600 might seem
impressive, especially if `advertised' on a proper 
scale (in linear scale the plateau level 
${\cal R}=1$ will be confused with 0,
and the curve $f(r\,|\,n_c=38)$
will appear very well separated  from $r=0$), 
and could easily convince non-experts that  
the searched-for signal exists. Nevertheless, 
confronted to such a result, there could be experts with strong
physically motivated priors who would still 
maintain their scepticism,
while others would hesitate. The reason is that in this 
domain of research prior knowledge is largely 
non-intersubjective. Even researchers who are members of 
the same experimental 
team do not usually share the same opinion, 
and the case of 38 events over an expectation 
of 20 is typical of those cases over which there could be disagreement:
disagreement which could cause the result to be left unpublished
for years, unless a charismatic and optimistic team spokesman 
persuaded his fellows to claim a discovery.

It is interesting to use Bayes' theorem in a reversed mode
to understand which kind of prior produces 
sceptical, hesitant and optimistic reactions. (We assume that 
researchers act in good faith and that they care about 
their reputation.) Above we have met two classes of priors: 
the uniform in $r$ and the uniform in $\log r$. One can 
easily imagine their effects, on the basis of 
the discussion concerning the example in Section \ref{inferring} 
(see Figs. \ref{fig:bayesflat}--\ref{fig:lowbgd}). 
But we do not think that there is a single 
physicist whose prior beliefs 
correspond exactly  to those of Eq. (\ref{eq:uniforme}) 
or (\ref{eq:Jeffreys}).  
It is much more reasonable to expect that someone 
would have a rough idea
of the order of magnitude of g.w. burst rate, 
provided that they exist at all.
 Now, an easy 
way to model an uncertain order of magnitude
is to think of a normal distribution in $\log r$, 
with most of the probability mass concentrated in some decades.  
The corresponding distribution of $r$ is called {\it lognormal}. 
Although this parametrization is, like any 
other, rough, 
it has some formal advantages which somehow reflect the  
prior knowledge of researchers: varying the two parameters
of the distribution,
one can choose the orders of magnitude of 
the value where the beliefs
are concentrated; 
the probability density function goes to zero as $r\rightarrow 0$
(in agreement with the working hypothesis that the  
searched-for signal does exist, and hence at a non-null rate);
the probability density function is defined for all 
positive values of $r$, thus capable of persuading  even initially 
very sceptical people to change their mind 
as soon as the  accumulated evidence  starts to 
produce a narrow peak in ${\cal R}$. 
When this situation of strong evidence 
is achieved, scientific
conclusions (summarized in  the final probability density function)
will not depend on the  details of the priors: 
All researchers will agree on the interpretation 
and the result will be considered objective 
(although, we repeat, it is only intersubjective).

\begin{figure}[!h]
\begin{center}
\begin{tabular}{|c|}\hline
\epsfig{file=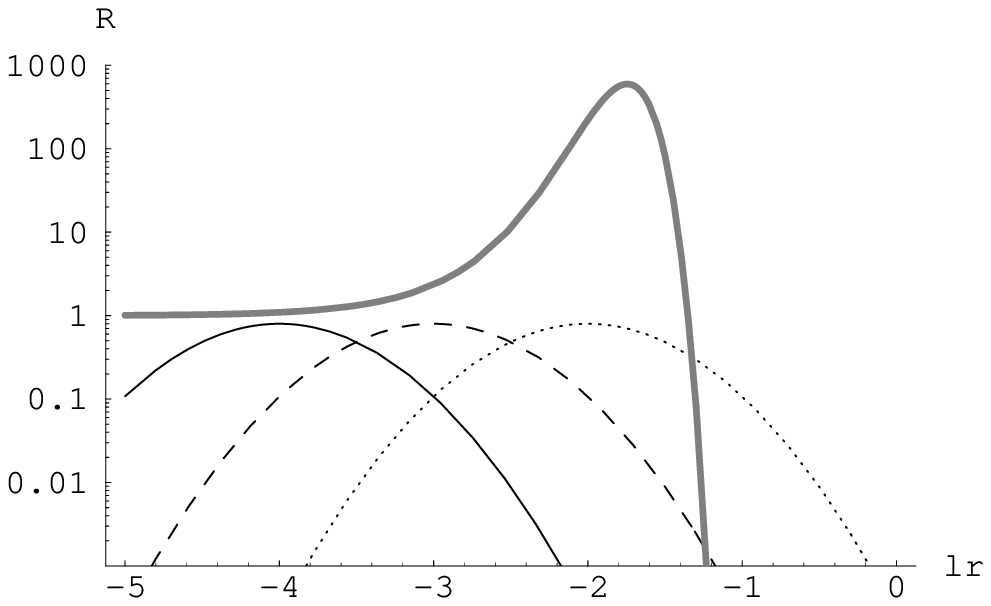,width=0.53\linewidth,clip=} \\ \hline
\epsfig{file=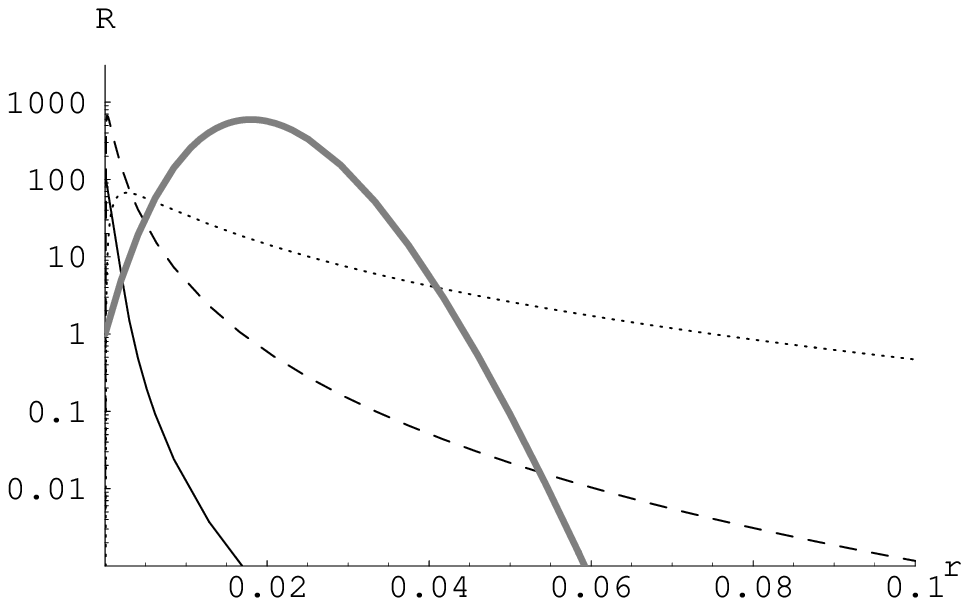,width=0.53\linewidth,clip=} \\ \hline
\end{tabular}
\end{center}
\caption{\small  Pessimistic (continuous), 
optimistic (dotted) priors 
plotted in different scales [$lr$ stands for $\log_{10}(r)$]. 
The dashed line represents 
an intermediate situation.
The grey curve is the  ${\cal R}$ 
function for 38 observed events out of 20 expected. }
%%Top: $\log_{10}(r)$
\label{fig:priors}
\end{figure}

\begin{figure}[!b]
\begin{center}
\begin{tabular}{|c|}\hline
\epsfig{file=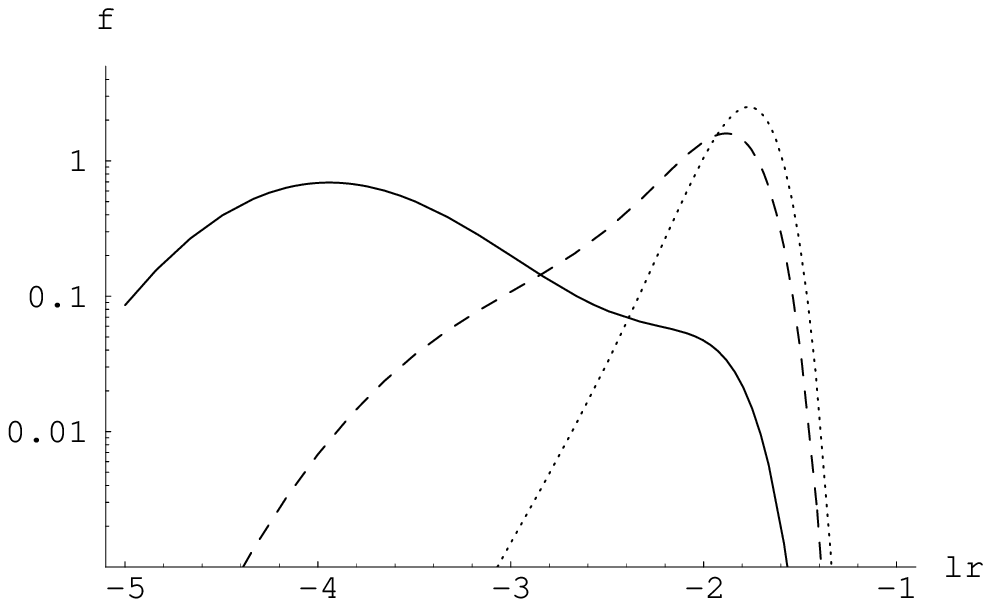,width=0.55\linewidth,clip=} \\ \hline
\epsfig{file=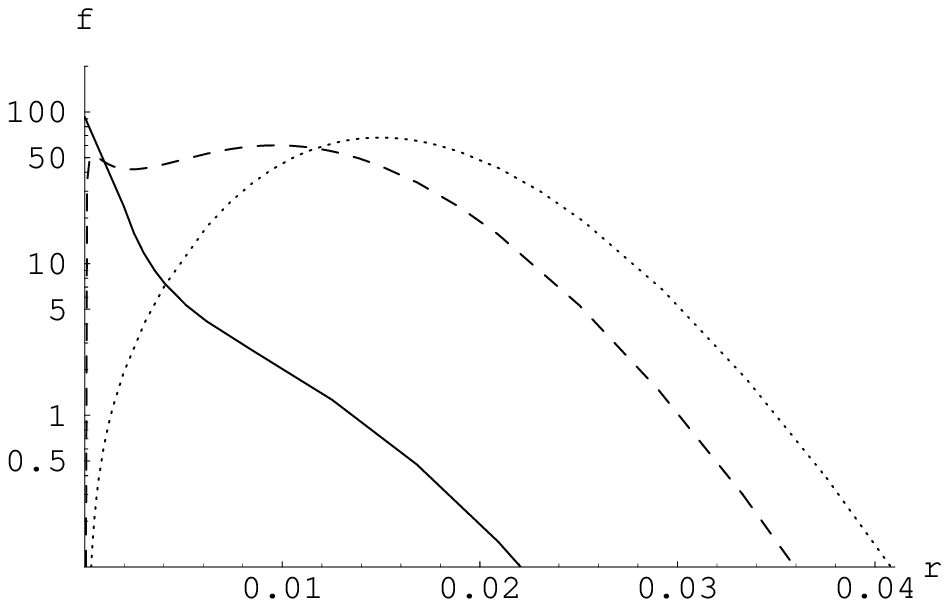,width=0.55\linewidth,clip=} \\ \hline
\epsfig{file=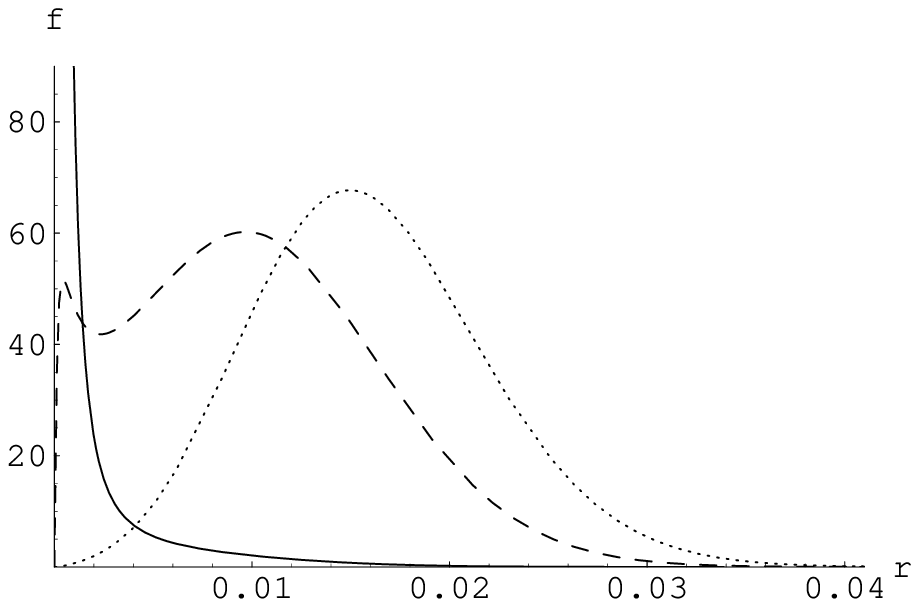,width=0.55\linewidth,clip=} \\ \hline
\end{tabular}
\end{center}
\vspace{2.0mm}
\caption{\small  Several representations of the final distributions
resulting from the three different 
priors of Fig.~\ref{fig:priors} and the evidence from 
38 observed
events out of 20 events expected from background (grey curve of 
 Fig.~\ref{fig:priors}). Note that the $f$ stands for the generic
symbol of p.d.f., but its 
mathematical function depends on the variable
via the Jacobian.}
\label{fig:finals}
\end{figure}

Considering the situation of 38 events observed
over an expected  background of 20 events, the prior knowledge
corresponding to the subsequent sceptical, hesitant 
and optimistic reactions can be 
modelled with a Gaussian in $lr = \log_{10}(r)$
having standard deviation 0.5 (i.e. half a decade) and averages
of $-4$, $-3$ and $-2$. 
\begin{table}
\caption{\small Dependence of initial and final probability for 
$r\ge 0.01$ bursts/day as a function of the  prior parameters (see text).}
\begin{center}
\begin{tabular}{|c|cc|cc|}\hline%\RequirePackage{a4}
Reaction  & \multicolumn{2}{|c|}{Prior parameters} & 
    \multicolumn{2}{|c|}{$P(r\ge 0.01\,\mbox{burst/day})$ } \\ \hline 
& $\mbox{E}[\log_{10}(r)]$ & $\sigma(\log_{10}(r))$ & prior  & final\\
sceptical & $-4$ & $0.5$ &  $3.2\cdot 10^{-5}$ & $0.9\,\%$ \\ 
hesitant  & $-3$ & $0.5$ &  $2.3\,\%$          & $50\,\%$ \\ 
optimistic & $-2$ & $0.5$ &  $50\,\%$           & $85\,\%$ \\ \hline
\end{tabular}
\end{center}
\label{tab:prob}
\end{table}
Table \ref{tab:prob} gives the parameters 
of the three priors, as well as the probabilities that $r$ is above 
0.01 bursts/day before and after the new experimental data. 
 Figures \ref{fig:priors} and \ref{fig:finals}  show
the modelled priors and the corresponding final distributions.
The figures are drawn with several 
scales because each way of representing them can help one to get a feeling
of what is going on. 

\subsection{Evolution of beliefs}\label{ss:evolution}
\begin{figure}[t]
\begin{center}
\begin{tabular}{|c|}\hline
\epsfig{file=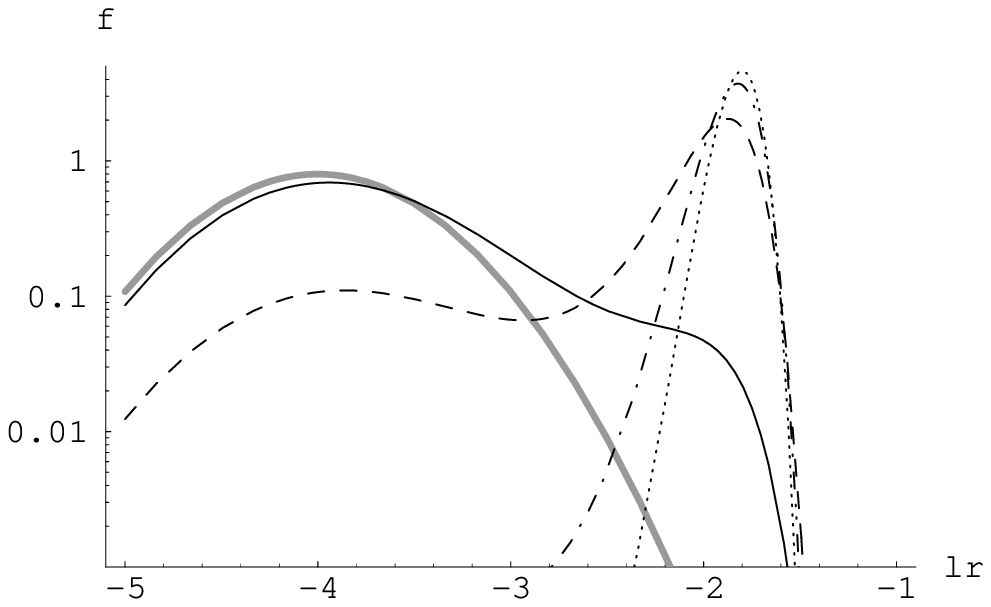,width=0.59\linewidth,clip=} \\ \hline
\epsfig{file=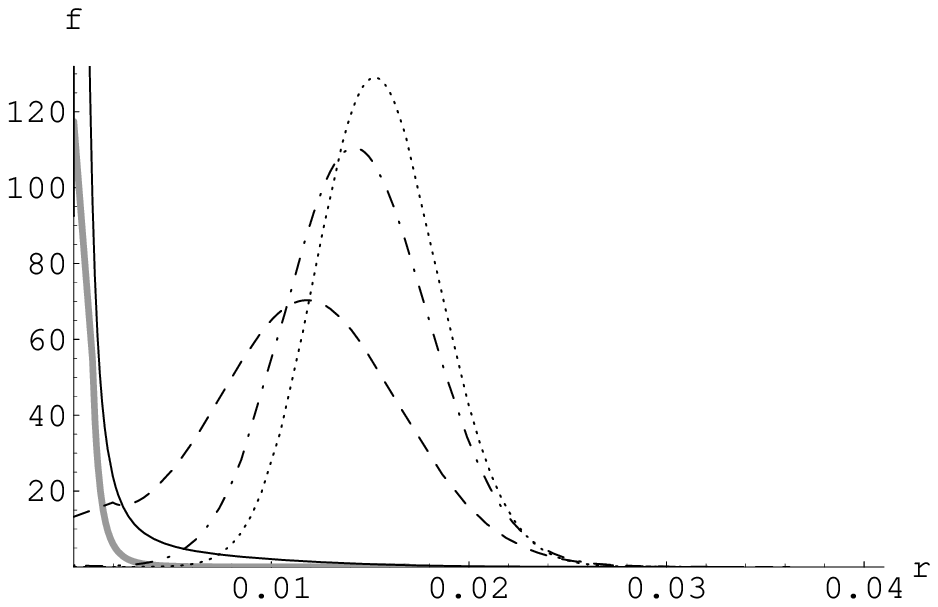,width=0.59\linewidth,clip=} \\ \hline
\end{tabular}
\end{center}
\caption{\small  Evolution of the p.d.f of a sceptical person 
(the grey curve is his/her initial prior) 
updated four times by independent evidence 
characterized by the same ${\cal R}$ function provided by 
38 observed events over an expected background of 20. Top
and bottom plots differ only by the scales.}
\label{fig:seq}
\end{figure}

We conclude this section with some remarks about the choice of the
priors used to illustrate this situation. 
First, it is clear that the 
chosen model is important, but what we want to show 
is the rough distribution of  beliefs. 
For example, although different parameters 
of the lognormal, or a different function of the probability, 
will produce numerical variations in the probability   
of the sceptical reaction, the qualitative conclusions will not 
change: Nobody possesses a psychological perception of probability 
which enables them to tell exactly whether their
intuitive probability is 0.5, 1 or 2\,$\%$.
What matters is that these probabilities are perceived as 
rather low and well separated from the range 
which characterizes hesitation ($\approx 40\mbox{--}60\,\%$) 
or almost certainty ($\gtrapprox 90\mbox{--}95\,\%$). 
Second, although we are not going to enter   
into the detail of trying to explain why the three different 
researchers have such different priors, it is important 
to understand that, since we have in mind real
researchers, priors are not simply  
abstract, aesthetic or philosophical ideas about the physical quantity.
They summarize a complex prior knowledge, based on
previous experimental observations
as well as on theoretical ideas related to this and  
other observables (we shall come back to this point in 
Section \ref{sec:does_exist}).
For example, looking at
the numbers in Table \ref{tab:prob} and the plots in 
Fig.~\ref{fig:finals} one could easily imagine that 
if the sceptical person was faced 
for a second time with independent evidence of similar
strength to that provided by the 38 observed events 
(which could again come from antenna data, 
but also from other astrophysical information), he/she could 
be now  in a situation similar to that of the hesitant person and the 
next time could be in the situation of the optimistic person. 
This evolution is illustrated in Fig.~\ref{fig:seq},
 which shows 
the p.d.f.  of the initially sceptical researcher
as he/she is faced four consecutive times with such rather
strong (and quantitatively identical -- clearly an academic exercise)
evidence. After the fourth occasion, he/she will
be strongly convinced that the rate is well above   
$10^{-2}$ bursts/day (see dotted curve of 
Fig.~\ref{fig:seq}). Asymptotically, when a large number
of pieces of evidence are in hand, the beliefs will be concentrated 
around the peak of the likelihood 
(i.e. $1.8\cdot 10^{-2}$ bursts/day), 
as the prior distribution becomes 
irrelevant.\footnote{Given the rough modelling of the sceptical prior
of this numerical example, one needs about 20 exposures to 
evidence of the kind considered, 
before the barycentre of the final distribution
reaches the simulated 
`true value' of $1.8\cdot 10^{-2}$ bursts/day. This is due to 
the rapidly decreasing tail of the log-normal distribution. 
It seems to us that, outside the order of 
magnitude considered more probable, the 
intuitive priors of  experienced physicists for 
this kind of frontier physics quantity have flatter 
tails. As a consequence, once the final distribution
has moved from the decades that one believed to be more
probable, the convergence to the true value becomes faster.}
 
%\section{Summarizing the result with a single number}
\section{Reporting a  result with an upper/lower bound}
Although the ${\cal R}$ function (which, we repeat, contains the 
same information as the likelihood function, but has the 
practical advantages we have illustrated)
represents the most complete 
and unbiased way of reporting the result, it might also
 be convenient to express with just one number
the result of a search which is considered 
by the researchers to be unfruitful.  
Before trying to give some recommendations, it is important
to start by saying that any attempt to find a precise prescription, 
with the hope that a single  number 
will summarize
the experimental information completely, is a false endeavour. Nowadays 
it is not difficult, in fact, to provide the complete function 
${\cal R}$, parametrized in some way,
no matter how complicated ${\cal R}$ might be. This parametrization 
could be posted on a web page, or sent on request, 
if it was too voluminous
for a published paper,
as might be the case for  multidimensional problems,  
or if several ${\cal R}$ functions 
are obtained, depending on different assumptions about systematic
effects or about the underlying 
phenomenology.\footnote{For example, the ZEUS Collaboration
has recently published polynomial parametrizations of log-likelihoods, 
which contain the same amount of information of ${\cal R}$, 
for each possible coupling of new contact interactions between
electrons and quarks~\cite{zeus}.}

If, anyway, one  
wants to report the result considered inconclusive
as a single 
number\footnote{As is well known,
in the case that ${\cal R}$ depends on two parameters,
like in neutrino oscillation analyses, one obtains a contour plot.}
having the meaning of a bound,
Fig.~\ref{fig:burloglog}
suggests that this number should be 
in the region of $r$  where ${\cal R}$
has a transition from 1 to 0. 
This value would then delimit (although roughly)   
 the region in which the $r$ 
values are most likely to be 
excluded from the region in which it 
is most likely that the true value lies.
 One may take, for example,  
the value of $r$ for which ${\cal R}(r)=5\,\%$ or 1\,\% of the 
insensitivity plateau\footnote{One could choose, alternatively,
a value corresponding to a percentage of the 
maximum of the likelihood, and hence of ${\cal R}$. As a convention, 
this could work as well, but: {\it a)} there is a problem 
of how to
handle local maxima of the likelihood in cases more 
complicated than the one under study; {\it b)} if the maximum of 
${\cal R }$ is high enough, the ${\cal R}$ function  corresponding
to the bound could be larger than 1. 
Note that, in the case of local maxima and minima 
of ${\cal R }$,  the condition ${\cal R}(r)=5\%$ or 1\%
yields multiple solutions. In this case it seems to us natural 
to choose the one farthest from the insensitivity region.} 
value (i.e. 0.05 or 0.01), 
or any other conventional number. What is important 
is not to call this value a bound at a given probability level
(or at a given confidence level -- the perception
of the result by the user will be the same!~\cite{maxent98}). 
This would be incorrect. 
In fact the ${\cal R}$ function is not sufficient by itself for
assessing a probabilistic
statement about the quantity of interest. 
 
If we had to suggest a possible convention for the 
upper bound, it
would be to choose $r_L$ such that ${\cal R}(r_L)=5\,\%$.
The advantage of this convention
 is that  it is easy to 
recover standard limits\footnote{In this particular
case the coincidence of the result obtained by  
the frequentistic prescription, the  ${\cal R}(r_L)=5\,\%$
convention, and the standard Bayesian result is due to a numerical 
effect due to the particular likelihood.}
 based on the rule ``upper limit equals 3'' 
(divided by the observation time)
when no events are observed.  
In this case, in fact, the likelihood 
is   
$f(n_c=0\,|\,r)=e^{-r\,T}$, and also 
${\cal R}(r\,;\,n_c=0) =e^{-r\,T}$. 
Moreover, a standard Bayesian
inference with $f_\circ(r)=k$ (see discussion in Section \ref{ss:pa})
produces the same\footnote{In the most general case, 
the 95\,\% probability bound limit will be 
different from that obtained by the ${\cal R}=5\,\%$ convention, 
and this is all right as the meaning of the two bounds is 
different (see previous footnote).} 
upper bound [Eq.~(\ref{eq:limit})]. 
 
We end this discussion about summarizing the 
objective result provided by the ${\cal R}$ function 
in one number with a few cautionary remarks. 
First, we repeat that this result should not be called 
a 95\,\% confidence level upper bound. 
Second,  
as can easily be seen from Fig.~\ref{fig:burloglog},
we do not think that it is worthwhile trying to define a 
conventional limit with the precision of a percentage level. 
Even defining the limit precisely, the 
expected statistical fluctuations of results from one 
experiment to another can easily change $r_L$ by 
$\approx 50\,\%$. Therefore, if one really wants
to quote a number for the upper limit, 
together with the ${\cal R}$ function,
one should simply state the order of 
magnitude\footnote{For a discussion about the significant 
digits of limits, see Sections 9.1.4 and
 9.3.5 of Ref.~\cite{dagocern}.}
 of $r_L$ obtained, for example, 
with the ${\cal R}=5\,\%$ convention.   

Finally, if one is interested in a limit having a probabilistic 
statement, one has to pass through the priors. 
In this case a possible way of producing conventional
probabilistic limits would be to use a uniform distribution, 
as discussed in Section \ref{ss:pa}. Upper bounds 
calculated as 95\,\% probability upper limits 
for the results of Fig.~\ref{fig:burloglog}
are given in Table \ref{tab:bounds} and are compared 
with the bounds obtained using the ${\cal R} = 5\,\%$ rule. 
We see that the results are very similar, 
if we remember that 
high accuracy in these bounds is not needed, as discussed 
above.\footnote{Another argument to understand our point is 
that the upper/lower limits should be considered  to 
belong to the same category of uncertainties, 
and not of true values.} 

\begin{table}[!h]
\caption{\small  Comparison of the upper bounds obtained 
on the rate $r$ using 
the ${\cal R} = 5\,\%$ rule, or evaluated 
as a 95\% probability upper limit
given by a Bayesian inference with uniform priors, with reference
to the example of Section \ref{ss:priors-free} 
(see Fig.~\ref{fig:burloglog}).
 The values 
of the upper bounds are rounded to remember that we do not
consider their exact value to be relevant (more digits are given 
within parentheses).}
\begin{center}
\begin{tabular}{lccccccc}\hline
%&&&&&&& \\
Observed number of events & 10  & 15 & 20  & 24  &  29  &  33  &  38 \\ 
%&&&&&&& \\ 
\hline 
%&&&&&&& \\  
95\% prob. ($10^{-2}$\,evts/day) 
     & 0.5          & 0.7          & 1.1          & 1.4          
& 2          & 2          & 3  \\
     & ({\it 0.51}) & ({\it 0.73}) & ({\it 1.06}) & ({\it 1.43}) 
& ({\it 1.96}) & ({\it 2.41}) & ({\it 2.98}) \\
%&&&&&&& \\  
${\cal R} = 5\,\%$ rule 
 ($10^{-2}$\,evts/day)  
     & 0.5          & 0.8          & 1.3          & 2  
& 3            & 4              & 5  \\ 
     & ({\it 0.54}) & ({\it 0.81}) & ({\it 1.30}) & ({\it 1.91})
& ({\it 2.90}) & ({\it 3.83})   & ({\it 5.13}) \\
%&&&&&&& \\ 
\hline
\end{tabular}
\end{center}
\label{tab:bounds}
\end{table}

\section{Does the searched-for process exist?}\label{sec:does_exist}
We have seen how to make the inference about the g.w. 
burst rate $r$, under the assumption that g.w. bursts exist. 
At this point some readers might have the objection
that they are  not interested in the values of the rate, 
but rather in whether g.w. bursts exist at all. 

It is quite well understood by 
scientists and philosophers that, while the observation
of a phenomenon
proves its existence, non-observation
does not prove non-existence
(the classic example that  
philosophers like for this reasoning is that of the black swan).
In our case, the problem is complicated by the fact 
that even the observations are not certain proof
of the existence. 
This is because, as long as some background 
events are  expected, we cannot be absolutely 
(mathematically) sure that the searched-for signal 
has been observed, no matter 
how many events are observed above those statistically expected
from background alone. This argument concerns not only 
the frontier problems we are dealing with in this paper, 
but all theoretical concepts, including true values of physical
quantities. 
So, to speak rigorously, we should only talk about beliefs. 
``Nevertheless, physics is objective, or at least that part 
of it that is at present well established, if we mean by 
`objective' that a rational individual cannot avoid believing it. 
\ldots The reason is that, after centuries of experimentation, 
theoretical work and successful predictions,
there is such a consistent network of beliefs, that it
has acquired the status
of an objective construction: one cannot mistrust 
one of the elements of the network without contradicting 
many others. Around this solid core 
of objective knowledge there are fuzzy borders which correspond to 
areas of present investigations,
where the level of intersubjectivity is still very low.''\cite{dagocern} 
As a consequence, it is not a question of proving
or disproving something (unless some impossible consequences have
been observed), but rather of how difficult it is 
to insert/remove something in/from what is considered 
to be the most likely network of beliefs. 

Applying these considerations to the case study, 
the answer to the question whether or not g.w. bursts exist
involves a complex knowledge of astrophysical and cosmological
facts and theories. 
As a consequence, we tend to believe that they could
exist until the experimental evidence is such that even 
the lowest conceivable rates of bursts carrying enough energy to pass
the effective energy threshold,  
evaluated by the best 
of our knowledge, are ruled out. On the other hand, 
we tend to believe that they have really been observed experimentally
when energy, rates and shapes of the signals match 
with the rest of the knowledge.\footnote{For example, if one 
observed a very high rate of very energetic bursts, which
seems incompatible with the possible sources in the Universe, 
most physicists would tend to believe that there was something wrong
with the experiment.}

\section{Dependence of the g.w. burst result on some systematic effects}
This last section is dedicated to systematic effects on the result.
Every experiment belonging to the class of inference 
that we are treating in this paper
has its own problematics. We consider here
only some of the effects which are more relevant for the case study 
we are dealing with, although some of the problems are common
to other experiments. As a general recommendation, 
we find very 
helpful the detailed study of the sources of uncertainty,
as e.g. listed in the ISO Guide~\cite{ISO}, and the use of conditional 
probability. For a general scheme for the evaluation of uncertainties
due to systematic effects, see Section 2.10.3 of Ref.~\cite{dagocern}. 

Returning to the inference of
the g.w. burst rate, we have considered so far a case study 
performed using realistic parameters,
but we have assumed 
ideal conditions concerning some aspects of the coincidence experiment. 
We will see below how the analysis strategy changes if we assume  
that the background is not perfectly known, or if it is not stationary; 
or if the physics process is not stationary. 
Before going into a discussion of these effects we need to 
consider the uncertainty about the optimal coincidence window
and about the minimum g.w. energy to
which the coincidence experiments are sensitive.

\subsection{Choice of the coincidence window}
\label{win}
The coincidence window  should be 
set  considering the physical behaviour of the sources, the distance
between the detectors and the detector characteristics, i.e. the
limited resolution introduced by the sampling time.
But other effects can influence the choice, such as
the noise that distorts the events or the fact that real
signals may have unexpected shapes.
So, in practice, the optimal coincidence 
window is usually chosen in order to 
maximize SNR, as a compromise between the demands for a 
reduced rate of accidental background on 
the one hand and for not missing
physical events on the other.
Therefore, the choice of the coincidence 
window involves unavoidably some arbitrariness,
and several values of the window can be envisaged, 
to cope with the possible 
assumptions about the signals looked for. As a conclusion,   
we do not think that there is just one way of 
reporting results, and the ${\cal R}$ values corresponding 
to different reasonable choices should be presented separately.  

\subsection{Uncertainty on  the minimum energy of g.w. bursts}
\label{sec:commenti}
At this point, we need to define  more precisely  
the quantity $r$ which is the subject 
 of the measurement (the measurand).
In fact, the case study assumed two parallel detectors 
responding to the same physical event which produced 
the burst of g.w.'s irradiating the Earth. 
This implies that $r$ is the rate of g.w. bursts with 
energy greater than the highest  energy threshold\footnote{We 
clarify that when we talk about energy, we really refer to
g.w. energy, and not to mechanical energy released to the antenna.}
of the two detectors. 
Calling the differential energy spectrum
 of g.w. bursts $\phi(E)$ ($=\mbox{d}r/\mbox{d}E$), we have 
\begin{equation}
r = r(E_{min}) \equiv \int_{E_{min}=E_{thr}^{max}}^\infty \phi(E)\,
\mbox{d}E\,,
\label{eq:def_r}
\end{equation}
which states that $r$ does, indeed, depend on
the minimal energy $E_{min}$ required for the burst 
to be detected. This minimal energy is related
to the maximum of the two threshold energies 
($E_{thr}^{max}$). 
Obviously, one will be interested in measuring the detailed 
$\phi(E)$, when the high sensitivity of future detectors 
will allow measurement of many burst candidates for different 
energy thresholds.  

It is easy to understand that the definition of the measurand given by 
Eq.~(\ref{eq:def_r})
does not correspond to what is actually detected. In fact there 
is not a one-to-one correspondence between the 
energy resulting from the filter ($E_m$) and   
 the energy of the burst. This is 
true for all kinds of measurements, 
with the only difference being that in the case of g.w. detection 
the spread of $E_m$ around the true energy $E$ can be rather large,
depending on SNR. In fact, 
the intrinsic physical reason for this spread 
is noise: the measured energy of the detected signal 
depends on the randomness of size and phase
of the noise at the moment of g.w. 
interaction~\cite{coinci,nostro1}.
A further source of spread is the performance of the 
filter used in the analysis, as shown in 
Ref. \cite{nostro2}. 

This kind of problem can be partially solved, at the expense
of a greater uncertainty about the measured quantity, 
if one is able to model the distortion of the spectrum 
$\phi(E)$ into $\phi(E_m)$. This can be done by mapping the
transition probability $E\rightarrow E_m$ with a p.d.f. $f(E_m\,|\,E)$, 
which, in a discrete approximation, can be thought of as a 
transfer matrix. The knowledge of this transfer matrix 
then allows $\phi(E_m)$ to be unfolded to infer
$\phi(E)$.\footnote{For examples of unfolding methods currently 
used in particle physics, when this kind of problem is 
encountered, see Refs. \cite{Blobel}, \cite{Zech_red} 
and \cite{unfold}.  
An elementary  introduction to the problem of 
unfolding methods, as well as of 
other simple methods, can be found in Ref.~\cite{Cowen}. 
More sophisticated methods for spectrum unfolding and, 
generally speaking, 
image reconstruction, 
are presented in Refs.~\cite{Weise} and \cite{Buck}, 
respectively.}  
However, unfolding g.w. burst energy 
spectra goes beyond
the purposes of this paper.
Therefore, hereafter it will be assumed
 that $r$ corresponds
to the definition of the measurand given in Eq. (\ref{eq:def_r}). 

\subsection{Non-stationarity of the signal}\label{flux}
Another assumption which entered in our previous considerations
is that the g.w. bursts  have a constant rate during the 
observation time. This assumption is consistent with the present 
status of knowledge and it leads researchers to model 
the  arrival time of the bursts 
with a Poisson process of constant intensity over the whole period of 
observation. Nevertheless, one can envisage analysing 
the data with the hope of finding evidence for g.w. bursts
in a short period, perhaps triggered by other independent observations
which happened in the same period.\footnote{Given the actual levels of
background in present-generation detectors, it is very unlikely that 
one would be persuaded that a genuine train of g.w. bursts 
had really arrived in a short observation
time, unless it happened to be a truly spectacular phenomenon. 
Nevertheless, one could `gate' the candidate events 
by other pieces of evidence coming from 
independent sources of information concerning something
that happened  in a narrow time
window. Such independent information could be related to 
optical, neutrino or $\gamma$-ray burst observations.} 
With this possibility in mind, it is preferable to 
analyse the data in subperiods, in order to exploit 
the potential of the information collected. 
The result over the full period of observation can be easily 
recovered by merging the partial results, as 
discussed in Section \ref{sec:bur}.
It is easy to prove that, in fact, the result over the full
period during which the noise has been stationary 
is exactly the same as can be evaluated by 
combining the subperiods, since
\begin{eqnarray}
{\cal R}=\Pi_i {\cal R}_i &=& 
e^{-r\,\sum_i T_i }\left(1+\frac{r}{r_{b}}\right)^{\sum_i n_{c_i}}\,
\nonumber \\
&=& e^{-r\,T}\left(1+\frac{r}{r_{b}}\right)^{n_{c}}\,,
\label{eq:combinazione}
\end{eqnarray}
where $T= \sum_i T_i$ and $n_c= \sum_i n_{c_i}$.
For this reason,  it is preferable 
to keep results on short periods of observations separate. 
  
\subsection{Uncertainty on the value of the  background rate}
\label{backg}
We have assumed that expected background rate 
is well known, as can be cross-checked using the off-timing
technique and estimation from individual background rates. 
Nevertheless, one may be in a state of uncertainty about $r_b$;
for example if the observation time is very small
at a given level of background (see discussion
below concerning the non-stationarity of background).
In this case $r_b$ will be
an uncertain number too and so will be characterized by a 
p.d.f. $f(r_b)$. One then has a  likelihood $f(n_c\,|\,r,r_b)$ 
for each possible value of $r_b$. The likelihood which 
takes into account all possible values of $r_b$, each weighted with its
degree of belief $f(r_b)$, is obtained
by the rules of probability, yielding
\begin{equation}
f(n_c\,|\,r) = \int\! f(n_c\,|\,r,r_b)\, f(r_b)\,\mbox{d}r_b\,.
\end{equation}
The case of a well-known background rate ($r_b=r_{b_\circ}$) is recovered 
when $f(r_b)\rightarrow \delta(r_b-r_{b_\circ})$,
where $\delta(\cdot)$ is the Dirac delta function. 
Note that $f(n_c\,|\,r)$ will no longer be a Poisson distribution, 
and therefore the expression of the ${\cal R}$ function will also 
be more complicated than that of Eq.~(\ref{eq:erre}),
although this is just a 
computational complication. 
This consideration 
 leads to the prediction that 
the frequency distribution of random coincidences
made over a long period of time, during which $r_b$ fluctuates,
can look quite different from a Poisson distribution.

A last remark concerns the meaning of $f(r_b)$. This function 
is meant to describe the uncertainty about the exact value 
of $r_b$ in a period which is considered to
be stable to the best of our knowledge, and not the measured 
frequency distribution of the background rate during 
a long period. If one knows that different subperiods 
each had a different value of $r_b$ (within the unavoidable uncertainty),
this information must be used in a different way, as will 
be shown in the next section.

\subsection{Non-stationarity of the noise}
\label{non}
One of the most important and unavoidable problems in 
coincidence experiments
is the non-stationarity of the noise.
In fact, the chance of detecting a g.w. 
of minimum energy $E_{min}$ depends not only on the 
filter threshold
but also on the level of noise. A high level of noise acts as
a high effective threshold for the  g.w. signals, 
and, therefore, the high rate of collected data 
with a filter threshold much lower than this effective threshold
contains no useful information for the coincidence analysis. 

One might envisage two possible strategies for the 
data-taking of the individual antenna.
\begin{itemize}
\item
 Fixed energy threshold, as, for example, used in Ref. 
\cite{bardonecchia}. The threshold is fixed at a constant
energy level independent of the detector noise.
It follows that the 
event rate due to the background varies 
according to the detector noise level. However, as said above, 
this option does not imply that the rate of events due to g.w.'s 
is constant. 
\item
Varying energy threshold, as, for example,  used in Ref.
\cite{coinci}. The threshold level is given in terms of SNR, 
i.e. the energy of the threshold 
varies according to the detector noise. 
The event rate 
due to the background remains constant, 
while that due to the
signals varies according to each threshold value 
(i.e. according to each
sensitivity level of the detector).
\end{itemize}
We are aware of the complexity of this problem,
and a full treatment of it goes beyond the purpose of this paper. 
However, we think that some general considerations can be made.  
The basic observation is that, as far as possible, 
in the final analysis 
one should try to keep 
the definition of the measurand fixed, 
as discussed above. If this goal is achieved (within the 
unavoidable uncertainties), although only in subperiods, 
it easy to  combine of the results referring to the same 
measurand by multiplying the ${\cal R}$ 
functions of each subperiod:
\beq
{\cal R}=\Pi_i {\cal R}_i=\Pi_i 
e^{-r\,T_i}\left(1+\frac{r}{r_{bi}}\right)^{n_{c_i}}\,.
\eeq
In contrast, results referring 
to different measurands, i.e. coincidences 
obtained at different effective thresholds, should be kept separate. 

As a consequence of the above considerations, the 
natural procedure seems to be
somewhat in between the two strategies outlined above.
During the data-taking it is preferable to vary the threshold 
setting in order to keep the SNR at its lowest possible value
and thus maximize the chance of detecting g.w. events. 
The large amount of background events which are collected in this
way can be reduced by applying a more sophisticated selection
before using the g.w. burst candidates 
for the coincidence procedure. 
Then, at the moment of the final analysis, the  
data should be reorganized 
according to the effective threshold   
of the less sensitive antenna (see Section \ref{sec:commenti}). 
This is equivalent 
to performing many experiments at different 
effective thresholds, the result of each of which 
should be presented separately. 

At this point, we think that
a very simple simulation could help to make our points clearer. 
Let us take a numerical  example by considering the case of
two periods over which the system was
stationary, with effective energy threshold $E_{th_i}$ ($i=1,2$).
We take each period of $T_i=1000$ days, for a total
observation time of 2000 days.\footnote{These
very long subperiods are chosen to give
 a sufficient number of 
coincidences when we consider only two data samples. 
Obviously, the same considerations hold if the subperiods 
have lengths of hours, as is more reasonable.}
In each time interval the background rate is indicated by $r_{b_i}$ and  
we simulate $n_{th_i}$ g.w. bursts,
plus a number of random coincidences exactly equal to the 
expectation value. Although the situation is obviously oversimplified, 
we think that it should help to make the general considerations more
easily comprehensible. 

\subsubsection{Consequences of analysing together data taken 
at different effective thresholds}
Let us consider $r_{b_1}=0.02$ events/day, yielding 
20 background events in $T_1$ at the effective 
energy threshold $E_{th_1}$. 
At this energy threshold we simulate 18 genuine g.w. bursts.
The result of this simulation is therefore $n_{c_1}=38$ 
observed coincidences. 

During the period  $T_2$ of a more noisy situation 
the threshold has been properly raised, to $E_{th_2} > E_{th_1}$,
in order to keep $r_{b_2}=0.02$ events/day.
Let us assume this result was obtained by doubling  
the threshold, i.e. 
$E_{th_2}/E_{th_1}=2$. 
On the other hand, the  number of 
coincidences due to g.w. bursts changes, 
as a certain fraction of the bursts will go below threshold.
To get the order of magnitude of the effect, 
let us take 
the number of sources within 
the sensitivity volume increase  
as  $d^{3}$,  where $d$ 
is the maximum Earth-source distance 
reachable at the chosen threshold. 
Instead, the energy of g.w.'s released in the 
antenna goes like
$d^{-2}$.
Thus, the rate of observable bursts is the ratio of the two energy
thresholds at the ${-3/2}$ power.
Then, during $T_2$ we get   
$n_{c_2}\approx 20 + 18 \times 2^{-3/2} \approx 26$.

The upper plot of Fig. \ref{fig:bursv} shows the situation: 
\begin{figure}[t]
\begin{center}
\begin{tabular}{|c|}\hline
\epsfig{file=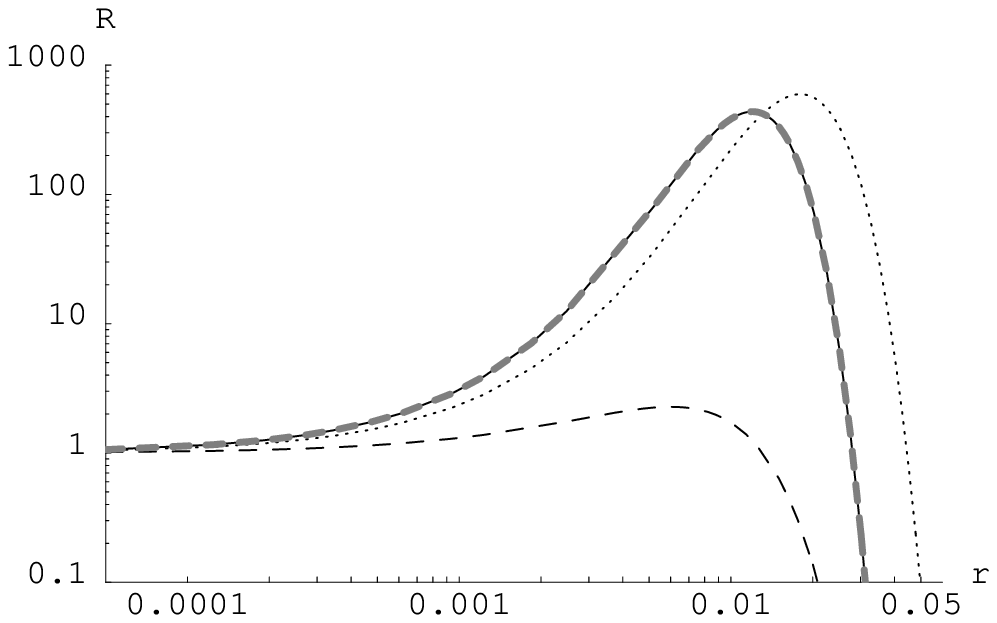,width=0.55\linewidth,clip=} \\ \hline
\epsfig{file=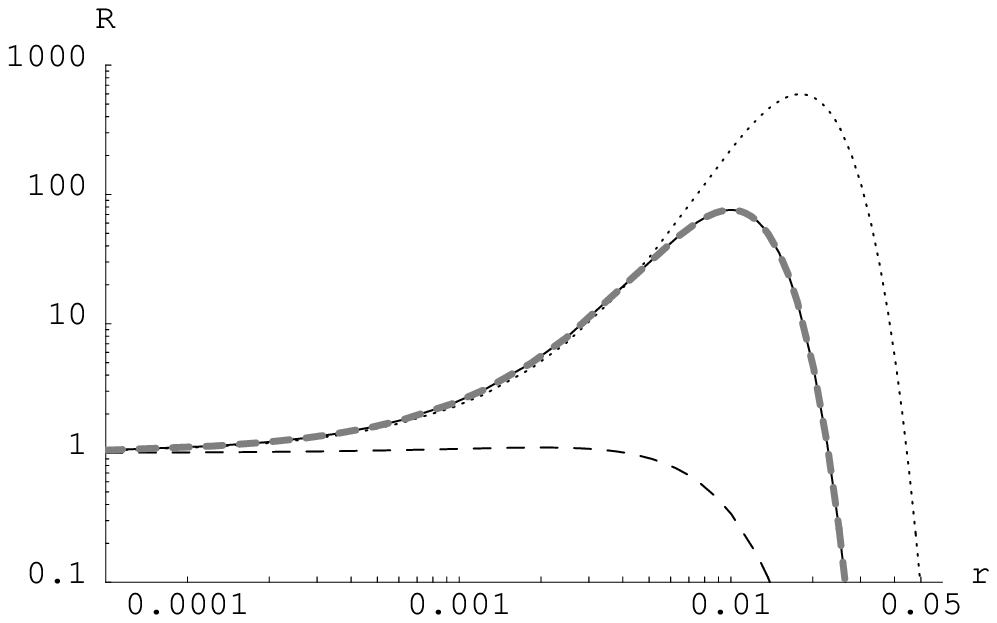,width=0.55\linewidth,clip=} \\ \hline
\end{tabular}
\end{center}
\caption{\small  Effect of na\"\i ve combination of data 
taken at different effective thresholds. In both plots
the dotted curve is the ${\cal R}$ function of the data taken in 
the low-noise period ($T_1$).
The black dashed curve is the ${\cal R}$ function of the 
data taken in the more noisy period ($T_2$) such that 
the threshold had to be varied by a factor of 2 (upper plot) 
or 5 (lower plot). The continuous and the grey dashed lines 
(overlapping) represent ${\cal R}_1\cdot {\cal R}_2$ 
and ${\cal R}_{AV}$ (see text).}
\label{fig:bursv}
\end{figure}
the dotted curve is the ${\cal R}_1$,
the black dashed curve is the ${\cal R}_2$ 
(its peak value is $\simeq 1.1$ ).
The  peak of ${\cal R}_1$ and the peak of ${\cal R}_2$ 
occur at different abscissae
($r_{m1}= 1.8\cdot 10^{-2}$ 
bursts/day, $r_{m2}= 0.6\cdot 10^{-2}$ bursts/day), as
expected due to the difference in the thresholds.
Since the two periods refer to different measurands,
the product ${\cal R}_1\cdot {\cal R}_2$ 
has  no inferential meaning.  Let us plot it, just to 
see what one would obtain by  making improper use of the combination 
rule given by Eq. (\ref{eq:bur_comb}).
Let us imagine also, for comparison, the full period analysed as a single
coincidence experiment. The corresponding  
${\cal R}$ function is indicated by ${\cal R}_{AV}$ and is obtained 
using a total number of   simulated coincidences
$n_c\approx 40+18+6 = 64$. 
Figure \ref{fig:bursv} shows that 
${\cal R}_1\cdot {\cal R}_2$ and ${\cal R}_{AV}$ coincide, 
but this does not justify the use of ${\cal R}_{AV}$, as 
we know that ${\cal R}_1\cdot {\cal R}_2$ is wrong too.
It is important to note that the position of the peak has moved
to a value strongly influenced 
by the data taken in the 
less sensitive period. The peak value is also reduced. 
Both these effects are a consequence of the mixing of different physics 
quantities.  
This effect can be shown more clearly by a new simulation in which the
effective threshold during $T_2$ is raised by a factor of 5
(bottom plot of Fig.~\ref{fig:bursv}). 
While the first period contains quite strong 
evidence in favour of g.w. bursts of energy $E > E_{th_1}$ and the 
second period provides 
a strong constraint for bursts of the higher 
energy $E_{th_2}$, the incorrect 
combinations mix up the two pieces of information, 
effectively spoiling both individual results. 

\subsubsection{Combination of data having the same effective threshold}
The results on the burst rate, with g.w.'s 
having a  minimum energy $E_{th_1}$, can 
only be obtained using the data collected during $T_1$. 
In contrast, information about g.w. bursts exceeding  $E_{th_2}$
can be obtained from both periods. The proper combination
of the two pieces of information is achieved by selecting 
the subsample of events taken during $T_1$ which have 
$E > E_{th_2}$. The rate of background events exceeding 
$E_{th2}$ can be evaluated by taking an exponential law 
relating threshold and rate, obtained assuming a
Gaussian noise for the amplitude\,\cite{nostro1}.
We then obtain  
$r_{b_1}^\prime=r_{b_1}\,e^{-E_{th_2}/E_{th_1}} \approx 
2.7\cdot 10^{-3}$. The number of observed events
in our simplified simulation is therefore 
$n_{c_1}^\prime=2.7 + 18\times 2^{-3/2}\approx 9$. 

\begin{figure}[t]
\begin{center}
\begin{tabular}{|c|}\hline
\epsfig{file=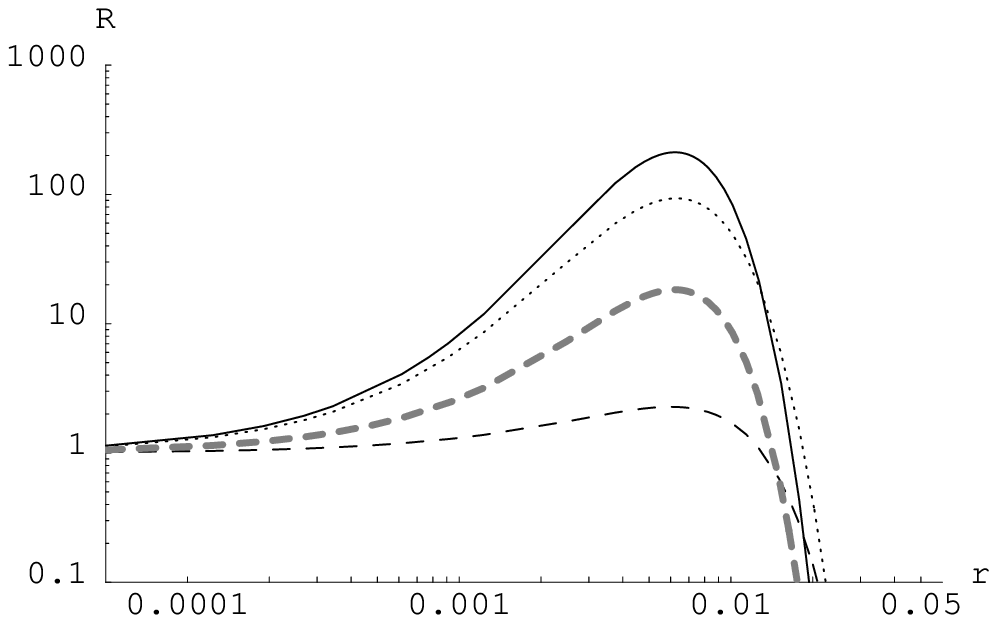,width=0.55\linewidth,clip=} \\ \hline
\epsfig{file=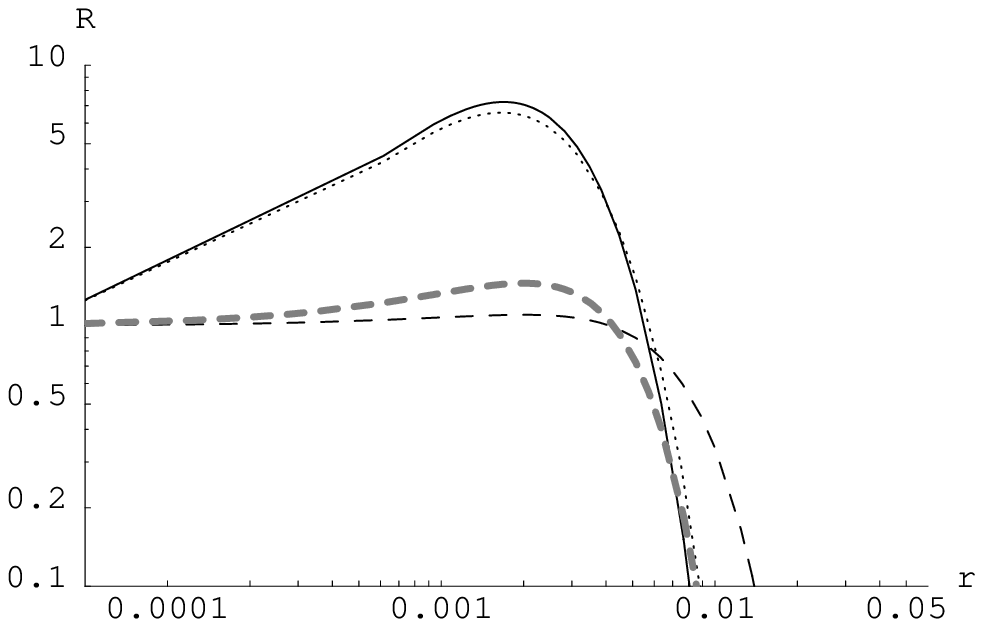,width=0.55\linewidth,clip=} \\ \hline
\end{tabular}
\end{center}
\caption{\small Combination
of data sets using the same effective energy threshold.
The dotted curves are the ${\cal R}$ functions of the data taken in
low noise period ($T_1$), selected to have an effective threshold
equal to the data taken in the more noisy period ($T_2$). 
The ratio of the selection threshold to the data taking threshold is 
a factor 2 (upper plot) and a factor 5 (lower plot).  
The continuous lines represent 
${\cal R}_{1\& 2}={\cal R}_1\cdot{\cal R}_2$, i.e. the 
correct combination of the two datasets. 
The grey dashed lines represent, instead, the result 
obtained by a na\"\i ve average of the two data sets.}  
\label{fig:bursf}
\end{figure}

The ${\cal R}$ functions relative to $E > E_{th_2}$
for the two periods are plotted in
the upper plot of 
Fig.~\ref{fig:bursf}: The dotted curve is  ${\cal R}_1$ 
and the black dashed  curve is ${\cal R}_2$.
The peaks of ${\cal R}_1$ and ${\cal R}_2$ are now 
both at $r_m= 0.6\cdot 10^{-2}$
bursts/day, as the data refer to the same
effective energy threshold.
The combined result is obtained by multiplying 
the two partial results 
(i.e. ${\cal R}_{1\& 2} = {\cal R}_1\,\cdot\, {\cal R}_2$)
and is shown by the continuous line in Fig.~\ref{fig:bursf}.
The evidence achieved by the combination of the two results is 
much better than was obtained in the good (i.e. low-noise)
period alone. 
This is a general result obtained in a natural way 
in the approach presented, and is in qualitative agreement 
with intuition. In fact, it is reasonable to think that, 
if data are analysed correctly, even a very noisy period, 
in which the detector is practically blind, should not 
spoil the evidence provided by the good period.
The overall evidence should increase as long as 
new data sets each containing a bit of information are 
added to the analysis. The formalization of these considerations
comes from the observation that 
${\cal R}$ has a constant value of 1 
for $r_b\rightarrow \infty$ [see 
footnote immediately after Eq. (\ref{eq:erre})].

The  ${\cal R}$ function obtained by
averaging the two periods, and indicated
again by  ${\cal R}_{AV}$, is now obtained using  
$r_{b_{AV}}=1.13\cdot 10^{-2}$ events/day and $n_c=35$ coincidences. 
The corresponding ${\cal R}_{AV}$
function is shown, for comparison, 
by the  grey dashed line in Fig.~\ref{fig:bursf}. 
This way of combining the data is unjustified, as it is not derived
from general rules of inference. Besides general arguments, the figure
shows that the na\"\i ve combination is less efficient 
 at using at best the evidence provided by the two sets
of data. 
As before, a new simulation in which the 
expected background rate during $T_2$ is five times that during $T_1$,
can illustrate more clearly this result. 
In fact, in the bottom plot of Fig.~\ref{fig:bursf}, 
one can now see that the noisy period provides only a very 
small piece of evidence; nevertheless, the 
correct combination of the two periods takes advantage
even of this very tiny piece of evidence, and 
the combined ${\cal R}_{1\& 2}$
has a peak slightly higher than ${\cal R}_1$. 
One can see that 
the na\"\i ve combination of the two periods, on the other hand, 
spoils the result 
obtained by the first period alone. This is obviously absurd: 
It is true that an infinitely noisy period brings no new 
information to the physical quantity of interest, but
neither should it
spoil the result achieved in the good period.

\section{Conclusions}
The problem of reporting the result about
the intensity of a Poisson process at the limit
of the detector sensitivity and in the presence of background
has been analysed from the perspective of 
probabilistic inference. This approach assumes that
probability is
related to the status of uncertainty and its
value classifies the plausibility of hypotheses
in the light of all available knowledge.  
We consider this approach the most general one
to draw probabilistic conclusions in conditions
of uncertainty, which is always the case when we want 
to infer the value of a physics quantity from 
experimental observations.

This approach is also 
known as Bayesian statistics because of the 
key role played by Bayes' theorem in updating 
probability in the light of new data. 
We have given arguments to show 
that Bayes' theorem is quite natural and produces 
results in qualitative agreement with intuition. 
That probabilistic conclusions depend also on
priors is natural too, although their presence 
tends to produce uneasiness in the practitioners.
This kind of `priors anxiety' can be overcome if one understands
their meaning and their role, which we have illustrated here 
with examples.  

We have shown that the contribution of the 
priors becomes irrelevant in routine cases, i.e. when the response
of the detector is very narrow around the true value. 
However, in frontier-science measurements, priors become 
crucial; so crucial that it is preferable to refrain from
providing probabilistic results. In this situation, 
the most objective way of reporting the result is to give 
directly likelihoods, or rescaled likelihoods in the form
of relative belief updating ratios 
(${\cal R}$ functions), described in this paper. 
The advantage of reporting ${\cal R}$ functions is that
they are easily perceived and  the combination of several 
experimental results can be achieved in the most 
efficient way. 

From the perspective illustrated in this paper, 
we consider a false problem 
that of finding a unique and objective 
prescription to calculate upper/lower limits (or contour curves, in 
the case of two-dimensional problems), 
which would summarize efficiently the result of the experiment
and would allow a consistent combination of results. 
Nowadays it is easy to provide
the complete ${\cal R}$ function, or several 
${\cal R}$ functions, depending on assumptions 
with regard to systematic effects.
Nevertheless, we understand that it can 
be practical to summarize 
the results with a number which roughly separates the region 
in which the experiment loses sensitivity (and ${\cal R}$ goes to 1) 
from the region practically ruled out by the data 
(${\cal R}\rightarrow 0$). 
This number can be based on a conventional 
value of the ${\cal R}$ function in the region of transition 
between 1  and 0. We have shown that similar numbers for the bounds can
be obtained using a standard Bayesian inference which uses 
a uniform prior. The upper/lower bounds
calculated 
in this latter  way can be interpreted as the  probabilistic limits 
that would be evaluated by the researchers sharing a 
positive attitude toward the possibility of the planned search.

The ideas illustrated in this paper have already been applied
to combine all pieces of evidence able to constrain the Higgs boson 
mass~\cite{higgs} and to the analysis of deep-inelastic scattering
events to search for new contact-type interactions between 
electrons and quarks~\cite{ci,zeus}. 
We have shown here that they are very useful in the analysis
of  gravitational wave bursts in coincidence experiments.
Indeed, the publication
of the results in terms of ${\cal R}$ functions 
for signals above a well-defined  
effective threshold (within unavoidable uncertainty)  
represents an efficient  way of taking advantage of
all possible pieces of evidence hidden in the data.

\section*{Acknowledgements}
It is a pleasure to thank P. Bonifazi, G. Degrassi, S. Frasca,    
G.V. Pallottino and  G. Pizzella for stimulating
discussions.  

\section*{Note added}
We would like to bring the attention of the reader to 
an interesting report~\cite{Eitel} which appeared  
while the present paper was going through the final 
editing procedures. 
Although much of the effort of the author has been 
dedicated to ``deduce correct confidence limits'', 
the report of Eitel shows for the first time 
(log-)likelihood functions of neutrino oscillation 
experiments as 3-D plots (Figs. 2, 5 and 13). 
Since offsets in log-likelihoods
are equivalent to factors in likelihoods, these results 
can be easily reinterpreted with the language developed in 
our paper: The asymptotic insensitivity region corresponds
to level 100 of Fig.~2 and level 0 of Fig.~13 (unfortunately, 
level $\approx 84$ is out of scale in Fig.~5). 
Moreover, the similarity between 
the curves of Fig.~14 of Ref.~\cite{Eitel}
and the ${\cal R}$ functions of our paper is self-evident. 
Indeed, these curves transmit the 
experimental result immediately and intuitively. 
Comparing Figs. 13 and 6 (and then extrapolating 
to Fig. 5, where the `flat ridge' is missing), 
one can realize how misleading 
the standard way of presenting neutrino oscillation results
as spots in the $\{\sin^22\theta,\,\Delta m^2\}$ plane can be.   
Figure 6 gives the impression  that LSND rules out all 
parameter space outside the spots. However, Fig.~14 
shows that LSND only rules out the parameter region which is
also excluded by KARMEN. Most of the complementary region 
is the region of insensitivity. In the boundary between 
these two regions (where KARMEN has already 
lost sensitivity), there is certainly a spot where there
is very high evidence (we assume no systematic effects 
have been overlooked), 
but this evidence cannot lead us to necessarely believe that the 
true values of $\sin^22\theta$ and $\Delta m^2$ are there, 
unless we have other reasons to believe it.

%\section*{References}

\end{document}